\documentclass{article}
\usepackage{arxiv}
\usepackage[utf8]{inputenc} 
\usepackage[T1]{fontenc}    
\usepackage{hyperref}       
\usepackage{url}         
\usepackage{booktabs}    
\usepackage{amsfonts}      
\usepackage{nicefrac}      
\usepackage{microtype}     
\usepackage{lipsum}
\usepackage{float}
\usepackage{graphicx}
\usepackage{caption}
\usepackage{subcaption}
\usepackage{titlesec}
\usepackage{multirow}
\usepackage{amsmath}
\usepackage{soul,color}

\usepackage[nodisplayskipstretch]{setspace}

\title{Inverse Design of composite Metal Oxide Optical Materials based on Deep Transfer Learning}
\author{
  Rongzhi Dong, Yabo Dan, Xiang Li \\
  School of mechanical engineering\\
  Guizhou university, \\
  Guiyang 550025, China\\
    \And
 Jianjun Hu *\\
 Department of Computer Science and Engineering\\
  University of South Carolina\\
  Columbia, SC 29201 \\
  \texttt{jianjunh@csc.sc.edu} \\
}
\titlespacing {\section}
{0pt}{5.5ex plus 1ex minus .2ex}{3.3ex plus .2ex}
\begin{document}
\maketitle
\begin{abstract}
Optical materials with special optical properties are widely used in a broad span of technologies, from computer displays to solar energy utilization leading to large dataset accumulated from years of extensive materials synthesis and optical characterization. Previously, machine learning models have been developed to predict the optical absorption spectrum from a materials characterization image or vice versa. Herein we propose TLOpt, a transfer learning based inverse optical materials design algorithm for suggesting material compositions with a desired target light absorption spectrum. Our approach is based on the combination of a deep neural network model and global optimization algorithms including a genetic algorithm and Bayesian optimization. A transfer learning strategy is employed to solve the small dataset issue in training the neural network predictor of optical absorption spectrum using the Magpie materials composition descriptor. Our extensive experiments show that our algorithm can inverse design the materials composition with stoichiometry with high accuracy. 

\end{abstract}

\section{Introduction}




Optical materials play a major role in imaging, communications, solar cell and sensor design, and the absorption spectra of meterials are the key research object for these applications. The optical properties of composite metal oxides are an interesting area of optical materials research because these properties may be very different from the properties of individual components. The structure of a composite material depends on its preparation process and the chemical properties of its constituent elements, which further affect its optical properties. Constituent elements of a composite material and the mole ratios among elements also have large impact on its structure and optical properties of the material. 

In order to study the optical properties of composite materials, first principles calculations such as Density Functional Theory (DFT) have been widely used \cite{sikam2019study,khalid2019synthesis}. Although first principles calculations are powerful, they are susceptible to the constraints of their excessive calculation cost, which limits the size of the material design space or the number of materials they can screen. To address this problem, machine learning (ML) has been increasingly applied to materials science fields, leading to the emergence of “materials informatics” \cite{rajan2005materials}, in which materials learning methods are developed to obtain prior knowledge and predictive models from known material dataset, and then predict complex material properties based on these models. In the past few years, ML has succeeded in predicting new features \cite{ward2017atomistic}, guiding chemical synthesis and discovering suitable compounds with target properties \cite{gomez2018automatic,popova2018deep,lu2018accelerated,collins2016materials}. While ML based material property prediction models can be used to screen known materials database to find candidates with expected properties, its performance is limited by the available materials, which are not developed for the target properties anyway. When ideal materials are not found in existing databases, discovering and synthesizing new materials with target properties is needed, which is usually based on the experience and knowledge of the researchers and expensive experiments. As materials that can be easily found have been found already and the scope of experimental exploration based on experience is narrow, new methods of material discovery is needed \cite{zunger2018inverse}, which promoted the development of the inverse material design approaches \cite{sanchez2018inverse}. 

Inverse design started in the field of alloy design \cite{ikeda1997new}, using genetic algorithm and molecular dynamics simulations to optimize the composition of multi-component alloys. This method received widespread attention in various fields once it was proposed, and now widely used in nanophotonic design \cite{molesky2018inverse,piggott2017fabrication,liu2018training,peurifoy2018nanophotonic,jiang2019simulator}, surface design \cite{liu2018generative,pestourie2018inverse,aharoni2018universal,jiang2019simulator} catalyst design \cite{freeze2019search}, catalyst design \cite{sanchez2018inverse,sanchez2017optimizing}, drug design \cite{popova2018deep} and materials design \cite{zunger2018inverse}. Inverse design of materials with desirable optical properties is of great significance in many industries such as solar cells, computer monitors, and optical microscopes\cite{doscher2014sunlight}. Inverse materials design can be regarded as an optimization problem, in which the desired materials characteristics are used as the optimization objectives. If the material is designed for optimizing only one attribute, it can be formulated as a single-objective optimization problem; if multiple materials attributes need to be optimized at the same time, it can be regarded as a multi-objective optimization problem. There are two major modules in a typical inverse design framework: one is the sampling module to guide the search in the design space in which a variety of optimization algorithms \cite{liao2020metaheuristic} can be used such as 
genetic algorithm (GA) \cite{qin2020genetic}, Bayesian optimization (BO) \cite{li2017bayesian}, particle swarm optimization (PSO) \cite{khadilkar2017inverse}, and differential evolution (DE) \cite{zhang2015inverse,bureerat2018inverse}. The other module is the forward property prediction model, which evaluates the performance of each design candidate, for which a set of commonly machine learning  algorithms have been used includinig support vector machines (SVM) \cite{wu2019learning}, random forest (RF) \cite{wirkert2016robust} and artificial neural network (ANN) \cite{sun2015artificial} etc. 

However, few studies have focused on inverse design of optical materials, especially inferring the possible compound formula of the materials only based on their absorption spectrum. The main reason of this phenomenon is lack of lage-scale optical characterizations dataset of  materials for model training. Previously, a metal oxide optical characterization dataset was published, and autoencoder algorithms for measured optical properties of metal oxides based on this dataset \cite{stein2019machine,stein2019synthesis} have been developed, which can map composite materials' characterization image patterns to its UV-vis absorption spectrum and vice versa. Comparing the band gap energy from the truth spectra to the predicted spectra, the root mean squared error and mean absolute error are 261 meV and 180 meV respectively, which are very small errors. However, it is not clear how to map the characterization images back to the composite material compositions. Yu et al. \cite{yu2013inverse} proposed a spectroscopic limited maximum efficiency metric, which can be used to guide the search of very thin film photovoltaic devices with high absorption. Inverse design of materials compositions have also been proposed by using generative adversarial networks \cite{dan2020generative}, which is mainly done by screening a large set of generated hypothetical materials.


Our work focuses on the inverse design problem of optical materials compositions with given target UV-vis absorption spectrum. In this study, we propose an algorithm to inverse design composite materials with a given target optical spectrum by combining artificial neural networks and genetic algorithms and Bayesian optimization. First, a large number of known composite material spectra are used to train a neural network model to predict the spectrum from the formula of a given composite material. The transfer learning method is used to train the spectrum prediction model of a specific set constituent elements with limited number of samples of varying stoichiometries.  Finally, the metal oxide material is designed inversely based on the given target optical absorption spectrum performance using GA and BO.

Our contributions can be summarized as follows:

\begin{itemize}
    \item We propose a transfer learning based neural network model for predicting optical absorption spectra from materials compositions. This approach helps to address the small data issue in materials property prediction.
    \item We develop an approach for inverse design of metal oxide material compositions for achieving a target optical absoroption spectrum using both genetic algorithms and Bayesian Optimization.
    \item We conduct extensive experiments and show that our proposed framework is capable to achieve good performance for target spectra.
\end{itemize}

The remainder of this paper is organized as follows. Section 2 focuses on the research framework, materials representation, and inverse design models of materials. Section 3 describes our experiments and highlights our inverse design performance. The last section concludes the paper.

\section{Materials and Methods}

\subsection{Problem setup and inverse design framework}

\begin{figure}[ht]
  \centering
  \includegraphics[width=0.8\linewidth]{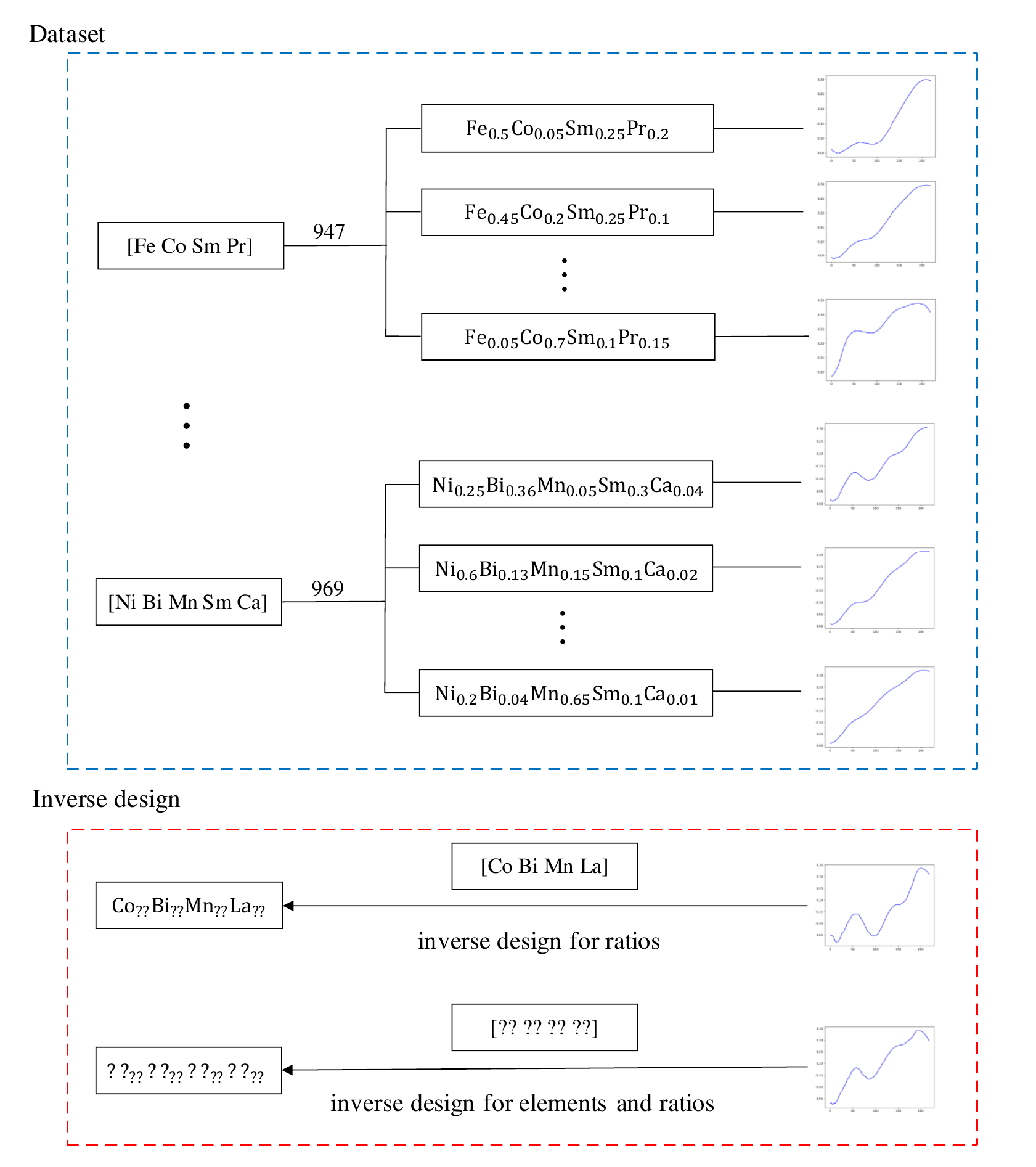}
  \caption{The inverse composite material design problem for achieving a target optical absorption spectrum.}
  \label{fig:problem}
\end{figure}

In our inverse design problem, the goal is to design the materials formula of a potential optical material that can achieve the given target light absorption spectrum. As shown in Figure \ref{fig:problem}, we have 554 different formula groups in the dataset, and each formula group consists of formulas with the same set of elements but different mole ratios. For each independent formula, there is a corresponding absorption spectrum. According to the composition of formulas, we define two versions of the inverse design problem: 1) the elements in the material are given, only the mole ratios need to be determined; 2) the elements are not specified in advance, we need to search both the elements and their mole ratios. 

\begin{figure}[ht]
    \centering
     \begin{subfigure}{1.0\textwidth}
        \includegraphics[width=\textwidth]{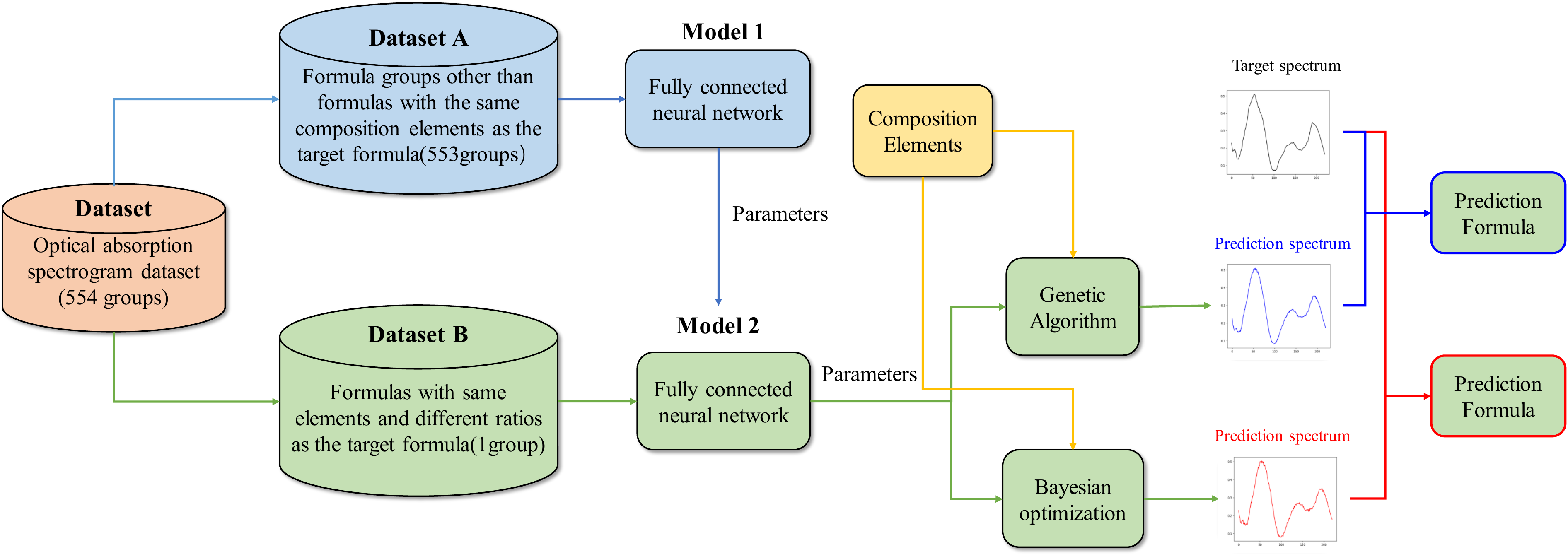}
        \caption{Transfer learning framework for computational inverse design of optical materials.}
    \end{subfigure} 
    \begin{subfigure}{1.0\textwidth}
        \includegraphics[width=\textwidth]{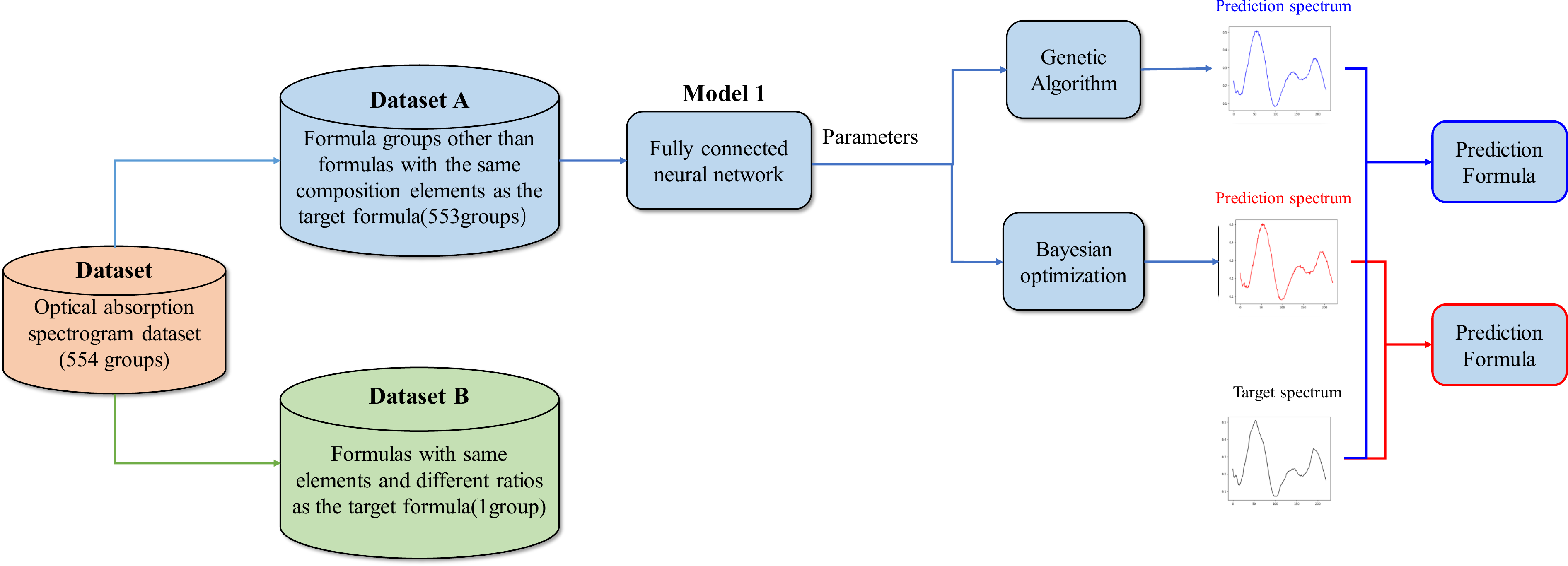}
        \caption{Framework without transfer learning for computational inverse design of optical materials.}
    \end{subfigure} 
    \caption{Frameworks for computational inverse design of optical materials.}
    \label{fig:framework}
\end{figure}

The main components of the framework are shown in Figure \ref{fig:framework}. We use deep neural network models, a type of machine learning model to learn the relationships between material composition and light absorption spectrum. Unlike previous work \cite{Stein2018Machine} which explores the relationship between characterization images of different optical material structures and their light absorption properties, we address the real-world need to inversely design suitable materials (in terms of  their materials composition) to achieve a specific target light absorption performance. In order to design materials according to performance specification and guide the discovery of new optical materials, we construct two inverse design models through the genetic algorithm and Bayesian optimization respectively to predict the corresponding material compound formula elements and their mole ratios based on the UV-vis absorption spectrum (220 items).

For inverse design with given elements, our framework is shown in Figure \ref{fig:framework} (a), which is composed of a fully connected neural network-based transfer learning model trained with Magpie features and global optimization based search model including a genetic algorithm and a Bayesian optimization. To divide the dataset, we randomly select the target formula from all 100,429 samples, and then set the formulas with same elements but different mole ratios as the target formula and corresponding spectra as Dataset B, while remaining 553 formula-spectrum groups are set to be Dataset A. Firstly, we use large amounts of known data (Dataset A) for initial training of the fully connected neural network model 1. Then we transfer parameters of model 1 to model 2 (with the same type as model 1), and use a small amount of sample data (Dataset B) to fine-tune the model. Finally, a genetic algorithm and Bayesian optimization method are used to inverse design materials that approximate the target optical properties through spectrum fitting. For inverse design without specifying the elements, although we also divide the dataset into A and B according to the known target, the fully connected neural network model is trained only by the Dataset A and without transfer learning(Figure \ref{fig:framework} (b)) to ensure fair comparison with transfer learning.

\subsection{Materials datasets}
\begin{figure}[h]
	\centering
	\begin{subfigure}{.45\textwidth}
		\includegraphics[width=\textwidth]{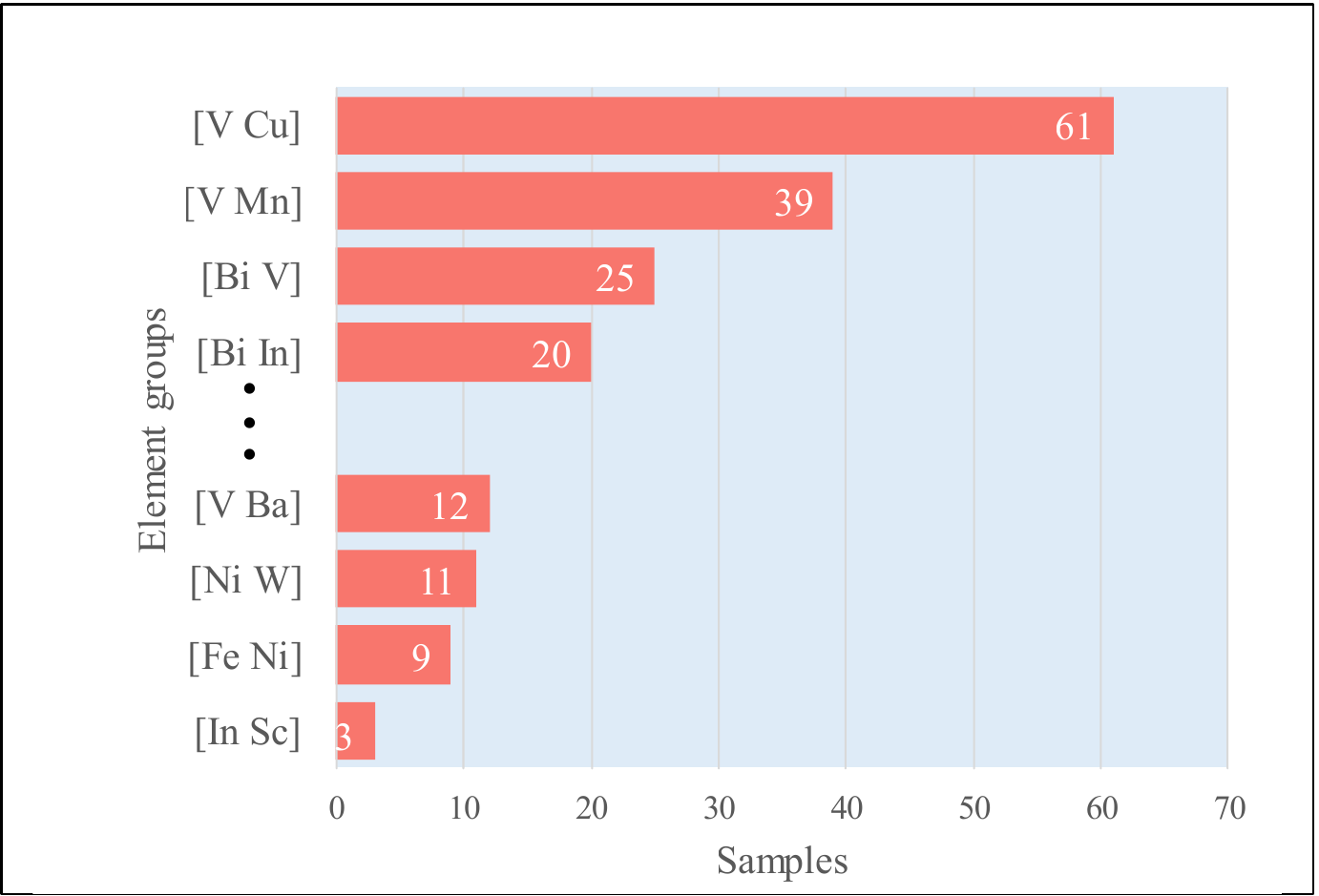}
		\caption{Number of samples for binary element groups.}
		\vspace{3pt}
	\end{subfigure}
	\begin{subfigure}{.45\textwidth}
		\includegraphics[width=\textwidth]{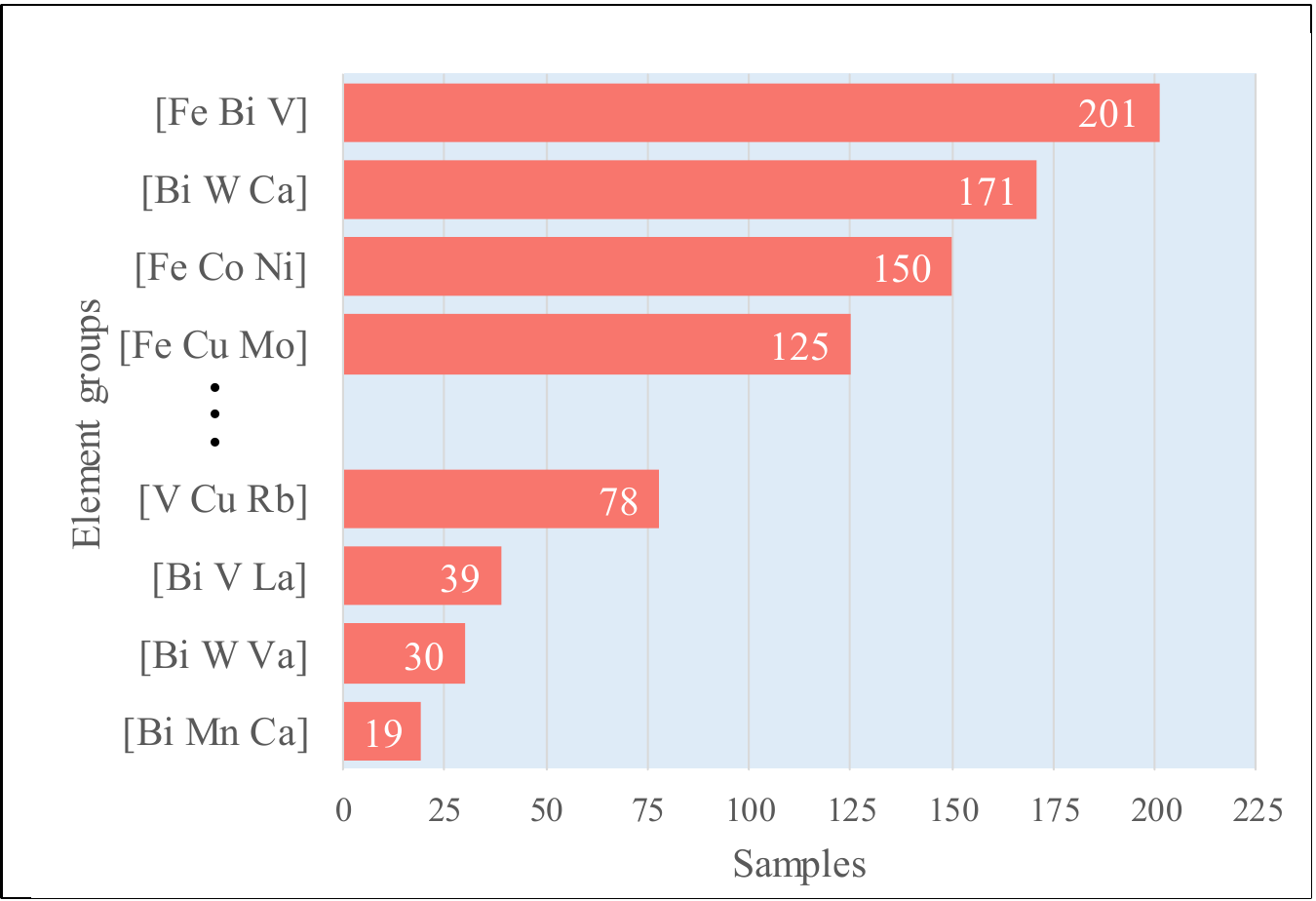}
		\caption{Number of samples for ternary element groups.}
		\vspace{3pt}
	\end{subfigure}
	\begin{subfigure}{.45\textwidth}
		\includegraphics[width=\textwidth]{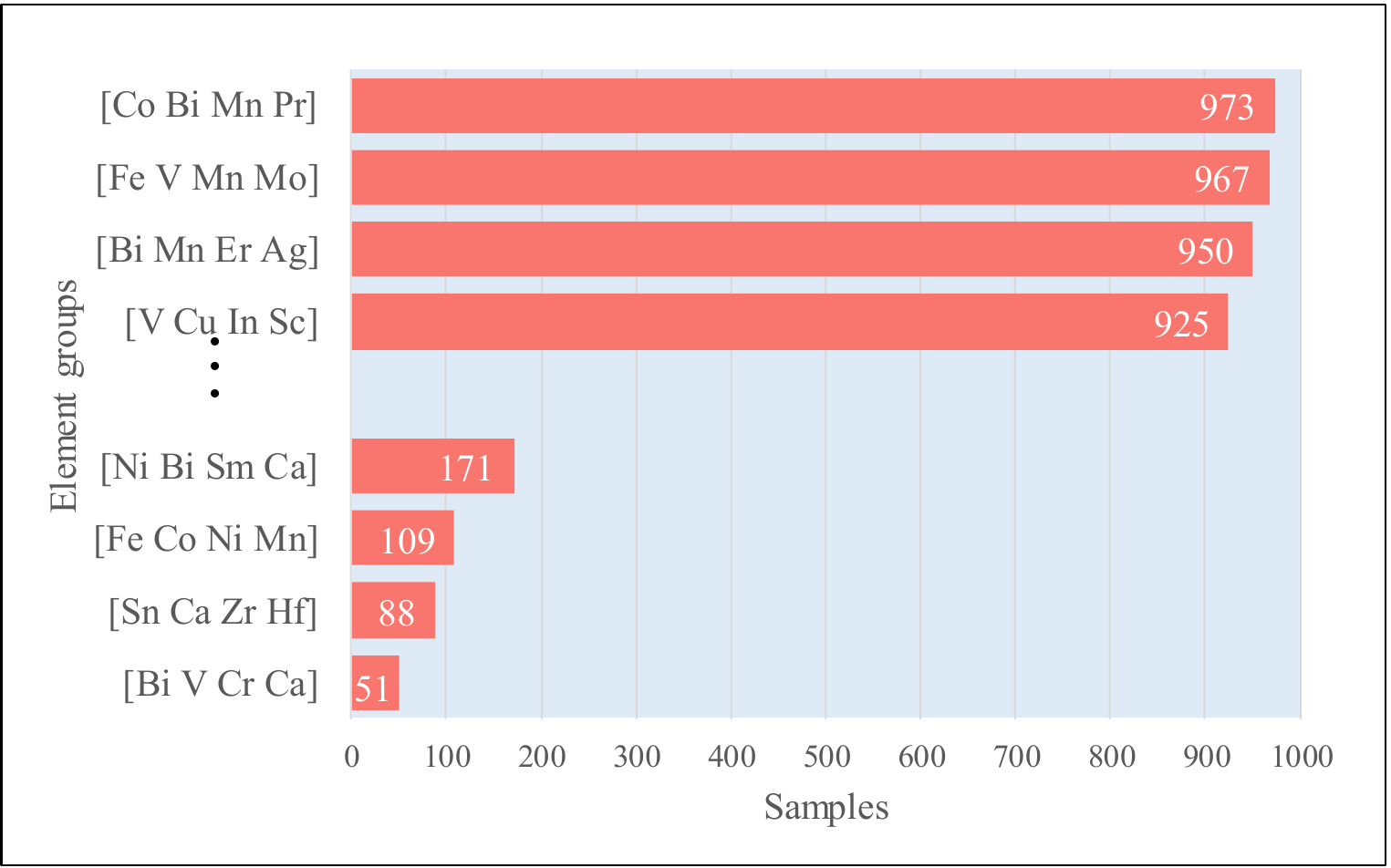}
		\caption{Number of samples for quaternary element groups.}
		\vspace{3pt}
	\end{subfigure}	
	\caption{Distribution of element groups in binary, ternary and quaternary compound materials.}
	\label{fig:samples}
\end{figure}

The material data used in this study are downloaded from references \cite{stein2019machine,stein2019synthesis}. This dataset contains sample images, UV-vis spectra and composition of a large set of metal oxide materials selected from the Materials Experiment and Analysis Database (MEAD) of the High-Throughput Experimentation (HTE) group at the Joint Center for Artificial Photosynthesis at Caltech. It is one of the largest publicly available scientific data set of cured metal oxide materials, which are synthesized by metal nitric acid salt with annealing. The absorption spectra are recorded using real-time scanning UV-vis dual-ball spectrometer while a flatbed scanner is used to obtain the sample images. This database contains a total of 178,994 molding material samples and their corresponding optical absorbance values at 220 energies between 1.32 to 3.2ev. The metal oxide samples contain various combinations of 1 to 5 cationic elements, as well as various inkjet printing and heat treatment parameters. These parameters are not used in the model described in this study. Since different preparation processes in the metal nitric acid salt produce many repetitive compound formulas, for materials with a fixed composition, we choose the average photoconductivity under different processes as its reference photoconductivity. After screening, we got a total of 100,429 samples composed of 42 different elements, which are then divided into 554 groups according to their constituent elements. Each group consists of a series of material formulas with the same composition elements and different mole ratios. Our dataset includes 2 groups of quinary compounds, 114 groups of quaternary compounds, 216 groups of ternary compounds, 181 groups of binary compounds, and 41 groups of simple substances. Figure \ref{fig:samples} shows the distribution of element groups in binary, ternary and quaternary compound materials.

\subsection{Materials representation}
We use the Materials Agnostic Platform for Informatics and Exploration (Magpie) composition descriptor \cite{ward2016general} to represent the materials in our dataset, which are calculated from properties of the atom elements in compound formulas to characterize materials. A Magpie descriptor vector is composed of a set of statistics of a selected element properties created by Ward \cite{ward2016general} and can be used for representing materials with any number of constituent elements. The set of properties is broad enough to capture a widely variety of physical and chemical properties which can be used to create predictive models of many material properties given only composition \cite{cao2019convolutional}. The physicochemical properties include stoichiometric properties (depending only on the ratios between elements), element properties (atomic number, atomic radius, melting temperature, etc.), electronic structure properties (valence electron number of s, p, d, and f layers) and ionic compound characteristics. To construct a Magpie feature, 22 weighted element attributes of the compound formula are calculated, and then the minimum, maximum, difference, average, variance and mode characteristics are calculated for each attribute. Finally, the material is characterized as a 132-dimensional data input. Here, the elemental properties are taken from the dataset available in the Wolfram programming language \cite{r1}.


\subsection{Model1 for spectrum prediction from composition: Initial training}

First, we select one target compound formula that needs to be inverse engineered. Except for the formulas in the same group as the target compound formula, we randomly choose 50 compound formulas for each group in Dataset A (select all formulas if less than 50), and then randomly divide them into the training set and the test set according to 70\%:30\% for performance  evaluation. The spectrum prediction model 1 used in this study is a fully connected multi-layer perceptron neural network model, and the model parameters are shown in Table \ref{table:table_dataset}. The batch size is designed by the user, in this article we selects 64 samples as a training batch.

\begin{table}[H] 
\begin{center}
\caption{Model parameters of the fully connected neural network network.}
\label{table:table_dataset}
\begin{tabular}{|l|l|l|}
\hline
\textbf{Layer} & \textbf{Input Shape} & \textbf{Output Shape} \\ \hline
Fc1            & {[}batch, input{]}   & {[}batch, 256{]}      \\ \hline
Fc2            & {[}batch, 256{]}     & {[}batch, 128{]}      \\ \hline
Fc3            & {[}batch, 128{]}     & {[}batch, 64{]}       \\ \hline
Fc4            & {[}batch, 64{]}      & {[}batch, 32{]}       \\ \hline
Fc5            & {[}batch, 32{]}      & {[}batch, 16{]}       \\ \hline
Fc6            & {[}batch, 16{]}      & {[}batch, 32{]}       \\ \hline
Fc7            & {[}batch, 32{]}      & {[}batch, 64{]}       \\ \hline
Fc8            & {[}batch, 64{]}      & {[}batch, 128{]}      \\ \hline
Fc9            & {[}batch, 128{]}     & {[}batch, 220{]}      \\ \hline
\end{tabular}
\end{center}
\end{table}

\subsection{Model2 for spectrum prediction trained by transfer learning}
In our inverse design framework, one key issue is that for a given element set, the number of training samples are too few. For example, there are only 51 samples in element set [Bi, V, Cr, Ca]. To address this issue, we use a machine learning strategy called transfer learning. Transfer learning \cite{zhuang2020comprehensive} is an algorithm to improve the performance of a target task in a target domain by exploiting some knowledge acquired when solving a source task in a source domain. Transfer learning has been widely used to address small dataset issue in machine learning. Usually, the source domain and the target domain or the source and target tasks are different. Transfer learning can reduce resource consumption and the time required for model training by just fine-tuning a trained model on the target task. In this study, the source domain is the compound formula groups other than the formulas with the same composition elements as the target formula in the previous step (Dataset A); and the target domain is a group of compound formulas with the same constituent elements but different element ratios as the target compound formula (Dataset B). Here both the source task and the target task are predicting light absorption performance. Our transfer learning strategy is to first train the neural network model on Dataset A and then import the parameters of this pretrained model 1 into the identical training model 2 for further training (fine-tuning). We randomly select 30 formulas in Dataset B and divide them into a training set and a validation set according to the ratio of 2:1, as the input to train model 2. The goal of fine-tuning the parameters for model 2 training is to make it more accurate to predict the composition of the target compound formula. 

\subsection{Genetic Algorithm}

A Genetic Algorithm (GA) is a global search method proposed by Holland \cite{holland1975adaptation} and Rechenberg \cite{huning1976evolutionsstrategie} inspired by the biological evolution in nature to search for optimal solutions. The algorithm transforms the optimization problem-solving process into a simulated evolution process with inheritance, mutation, selection, and crossover of chromosomal genes in biological evolution through mathematical methods and computer simulation operations. When solving more complex combinatorial optimization problems, compared to conventional optimization algorithms, better optimization results can usually be obtained. Genetic algorithms have been widely used in combinatorial optimization, machine learning, signal processing and adaptive control \cite{srinivas1994genetic}. In the context of materials science, genetic algorithms have been widely used in crystal structure prediction \cite{pakhnova2020search}, inverse materials design \cite{jennings2019genetic}, and materials property prediction \cite{shariati2020prediction}. 

\begin{figure}[ht]
	\centering
	\begin{subfigure}{.49\textwidth}
		\includegraphics[width=\textwidth]{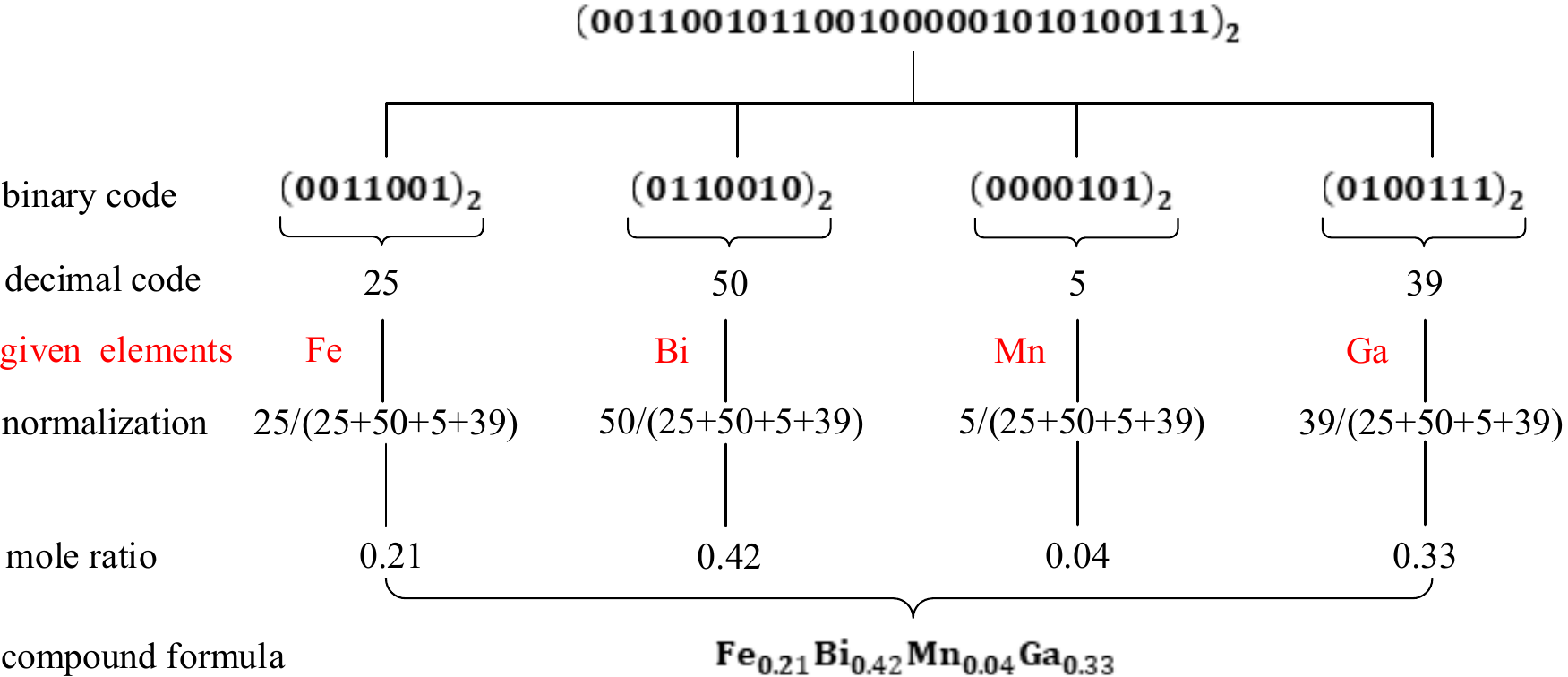}
		\caption{Redundant decoding approach.}
		\vspace{3pt}
	\end{subfigure}
	\begin{subfigure}{.49\textwidth}
		\includegraphics[width=\textwidth]{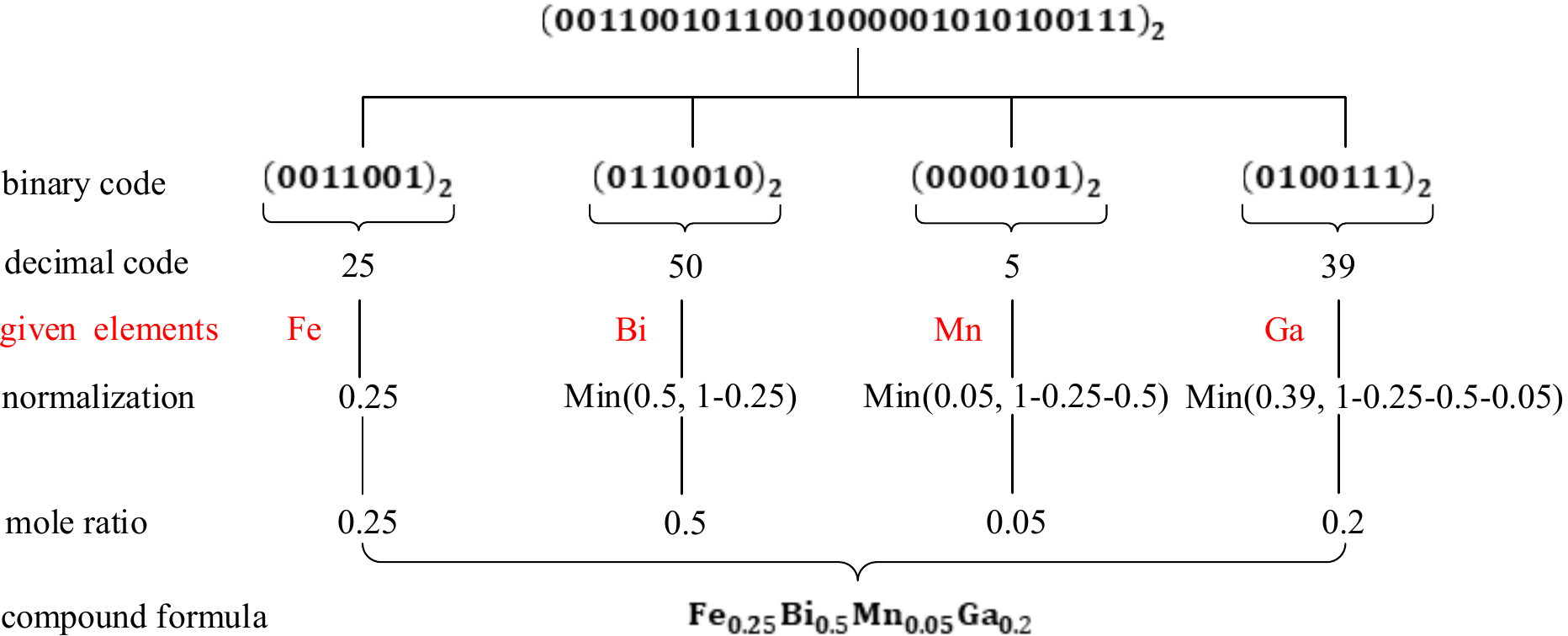}
		\caption{Unique decoding approach.}
		\vspace{3pt}
	\end{subfigure}
	\caption{Genetic encoding approaches for mole ratios with a given element set.}
	\label{fig:encoding1}
\end{figure}

As Gobin and Schuth \cite{Gobin2008On} clarified, the way that materials are presented in the genome may have a significant impact on the performance of evolutionary algorithms. In this study, each material is mathematically represented by a $n \times 7$ bit binary sequence  (its chromosome or genotype) representing n mole ratios of the n elements in the material composition (Figure \ref{fig:encoding1}). Each element's mole ratio in the compound formula is represented by a seven-bit binary string. Decoding the binary strings into decimal values, the mole ratios of the elements can be obtained, and the material composition can be determined. The binary sequences of the population can then be mutated and crossovered by specific genetic operators.

Two decoding strategies have been proposed in our study. In the first decoding approach (Figure \ref{fig:encoding1}(a)), the 7-bit string that encodes the mole ratio for each element is first decoded into a decimal value, and then each of them will be divided by the sum of these decimal values to ensure that the final sum of the decoded mole ratios to be 1. In the second  unique decoding approach as shown in Figure \ref{fig:encoding1}(b), the decoding process is as follows: 1) first all 7-binary strings are decoded into decimal value and then converted into a value between [0,1]; 2) the first ratio $r_1$ will be assigned as the mole ratio of the first element, for remaining ratio $r_i$, we first compare it with $1-\sum_{i=1}^{i-1}{r}_{i}$, then we choose the smaller one as the mole ratio of element $i$. By converting binary encoding to decimal values, the mole ratios of the elements can be obtained, and the material composition can be determined. Compared to the first decoding approach, it has the benefit that each mole ratio vector corresponds to a unique genetic binary string while the first approach allows redundant encodings for the same mole ratio vector.  


\begin{figure}[h]
  \centering
  \includegraphics[width=0.95\linewidth]{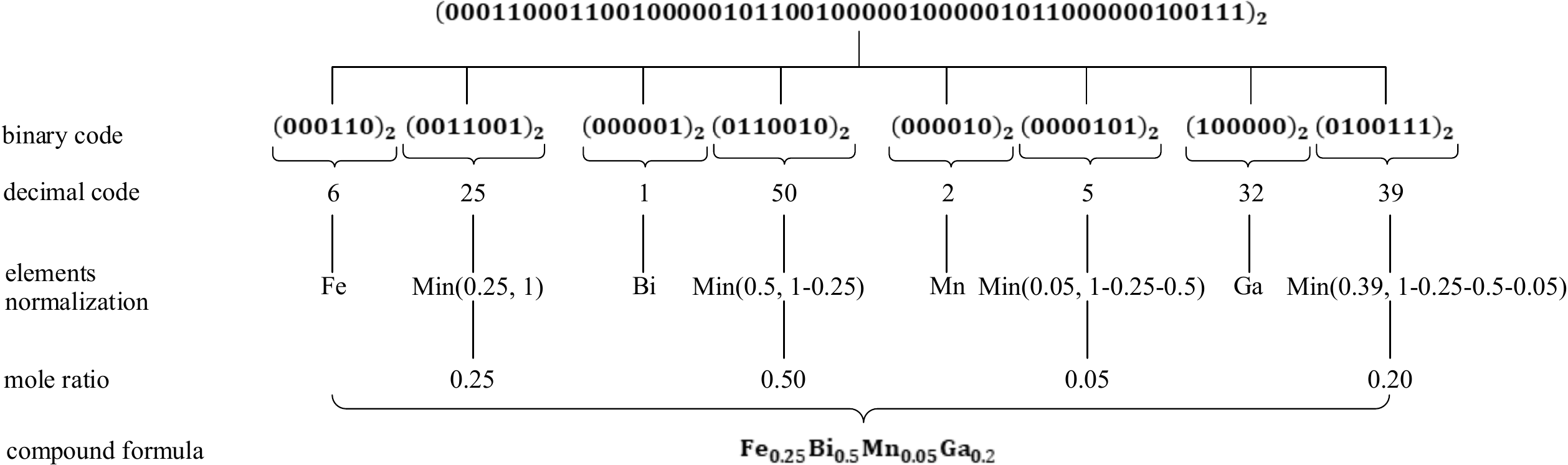}
  \caption{Genetic encoding for inverse design without specifying the element set}
  \label{fig:encoding2}
\end{figure}

For the inverse design problem without specifying the element set, the encoding of the GA is shown in Figure \ref{fig:encoding2}. The main difference is that the chromosome now has a block for encoding the $n$ elements, where $n$ should be specified by the user.

\begin{figure}[h]
  \centering
  \includegraphics[width=0.8\linewidth]{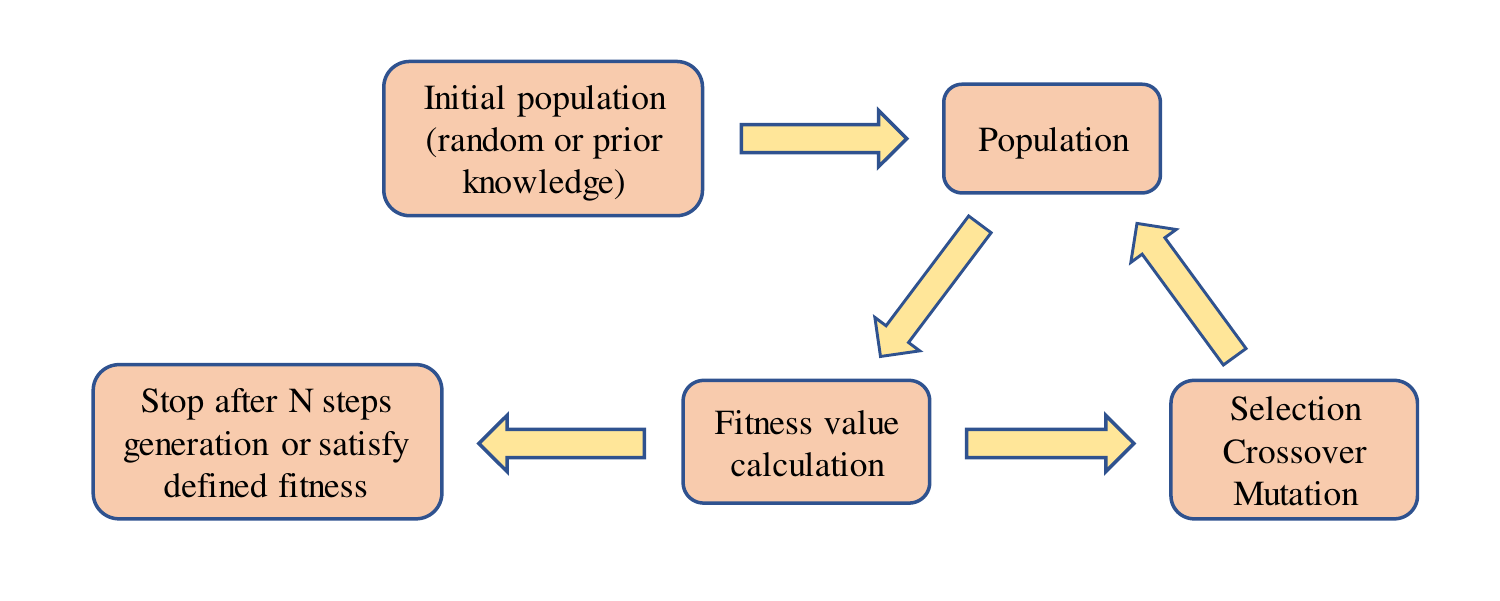}
  \caption{Basic steps of a genetic algorithm}
  \label{fig:GA}
\end{figure}

The initial population of individuals is usually randomly generated, but it can also seeded with suitable known materials. The generations are mutated to generate the next population in an iterative manner. The fitness of each individual in the population is evaluated by a fitness or objective function (here it is the MAE distance between the predicted spectrum for a material composition and the target spectrum). The fitness function may also include undesirable characteristics as constraints that need to be avoided. Then following the idea of survival of the fittest principle, a set of fitter materials will be selected from the current population for breeding via mutation or crossover operations on their genomes to generate a new generation of population.  This iterative cycle continues until the maximum number of generations is produced, or some members of the population have characteristics that reach the expected target. Figure \ref{fig:GA} summarizes the basic steps of the evolution process of a GA. The hyper-parameters of a GA include the material genome encoding length, the population size, the mutation and crossover rate, and the number of generations.

In this study, the genetic algorithm is used as one of the global search algorithms for inverse material design with a target light absorption performance. The evolution process starts with compound formulas with the same constituent elements but different element ratios as the target compound formula. The parameters of the genetic algorithm in this study are: gene length: 13 (6 for elements and 7 for ratios), initial population size: 500, crossover probability: 0.5, mutation probability: 0.5, generation: 100.

\subsection{Bayesian optimization model}

\begin{figure}[h!]
  \centering
  \includegraphics[width=0.5\linewidth]{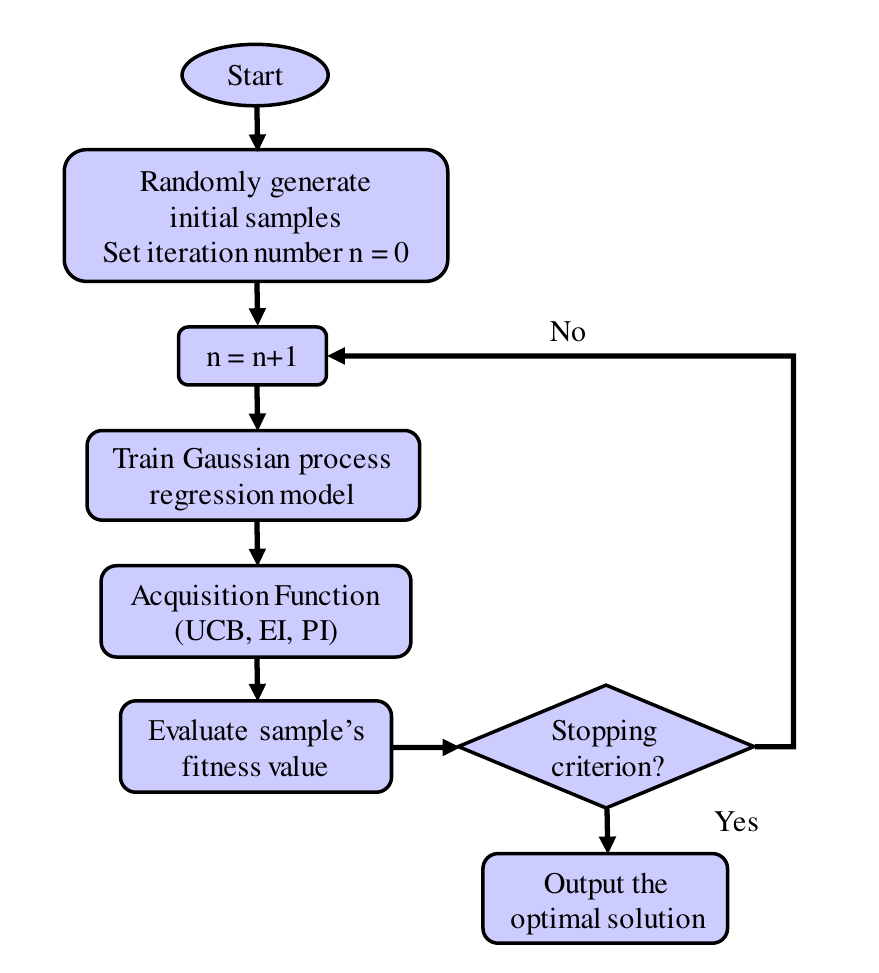}
  \caption{The flowchart of Bayesian optimization algorithm}
  \label{fig:BO-model}
\end{figure}

The Bayesian optimization (BO) method was proposed by \cite{kushner1964new}. Jones et al. \cite{jones1998efficient} introduced an effective global optimization (EGO) method and extended the BO technique. This method has become very popular and well-known in engineering and is now widely used in the design of time-consuming experiments, aimed at reducing experiment costs. Application of BO in machine learning mainly focuses on adjusting the hyperparameters of computationally expensive machine learning models \cite{snoek2012practical}. In this study, we use BO methods to build predictive models for the potential relationships between design variables of the materials and their properties, and then use decision theory to suggest which design is most valuable. BO finds the candidate solutions that minimize the objective function by establishing a substitution function (e.g. Gaussian process model) based on the evaluation results of the objective function. The Bayesian method is different from random or grid search in that it exploits evaluated sampling points to build a surrogate model which not only predict the objective values but also related uncertainty, which allow it to achieve automated balance of exploitation and exploration. The schematic diagram of BO algorithm is shown in Figure \ref{fig:BO-model}. 

BO requires several initial sample points, and through Gaussian process regression (assuming that the optimization variables conform to the joint Gaussian distribution), the posterior probability distribution of first n points is calculated to obtain expectation, mean and variance. The mean represents the final expected effect of this point, the larger the mean, the greater the final index of the model. The variance represents the uncertainty of the effect of this point, the larger the variance, the more uncertain the value of this point and worth exploring. Therefore, the first step in BO is to implement the Gaussian process regression algorithm. 

Another important step is to balance exploitation and exploration. For exploitation, points close to the known points need to be selected as the reference points for the next iteration, that is, excavate points around the known points, the distribution of points will appear in a dense area, which is easy to enter the local maximum. For exploration, points far away from the known points need to be chosen as the reference points for the next iteration, and make the distribution of points as even as possible to explore the unknown area. Sampling points with large mean value can be selected for exploitation and samples with large variance can be selected for exploration. To control the ratio of exploitation and exploration, the acquisition function needs to be defined. The simplest acquisition function is Upper confidence bound algorithm (UCB) which equals the mean plus k times the variance. Where k is the adjustment parameter, which can be intuitively understood as the upper confidence boundary. More complex acquisition functions include expected improvement, entropy search, and so on.

In this study, BO algorithm is used as one of the optimization algorithms for material light absorption performance inverse design. The following configuration parameters are set: the initial population size is 500, the number of generations is 100, and the UCB acquisition function is used to achieve the balance of exploration and exploitation.

\subsection{Model evaluation and experiment environment }

This study uses fully connected neural networks to construct training spectrum prediction models. We choose average absolute error (MAE), root mean square error (RMSE), and coefficient of determination (R$^2$) as the evaluation criteria of these models. MAE is used to reflect the actual situation of the predicted value error, RMSE is used to measure the difference between the predicted value and the true value, R$^2$ is used to indicate the degree of fit between the predicted value and the true value. The specific calculation formulas are as follows:


\begin{equation}MAE=\frac{1}{m} \sum_{i=1}^{m}\left|\vec{y}_{i}-\hat{\vec{y}}_{i}\right|\end{equation}

\begin{equation}RMSE=\sqrt{\frac{1}{m} \sum_{i=1}^{m}\left(\vec{y}_{i}-\hat{\vec{y}}_{i}\right)^{2}}\end{equation}

\begin{equation}R^{2}=1-\frac{\sum_{i=1}^{m}\left(\vec{y}_{i}-\hat{\vec{y}}_{i}\right)^{2}}{\sum_{i=1}^{m}\left(\vec{y}_{i}-\bar{\vec{y}}\right)^{2}}\end{equation}

where m is the number of samples, $\vec{y}_i$ and $\hat{\vec{y}}_i$ are the true and predicted values of the $i$ sample label (the spectrum of formula $i$), $\bar{\vec{y}}$ is the average of the $m$ sample real labels.
All calculations were performed on a Dell Server workstation equipped with an Intel Xeon W-2123 @3.70 GHz CPU, 16 GB RAM, a Nvidia GTX1080Ti GPU with 12 GB dedicated GPU memory. Software used was Python version 3.6.4, Keras version 2.2.0, and TensorFlow version 1.14.0. The random train-test split is 70\% for training and 30\% for testing.

\section{Results and discussion}

\subsection{Prediction performance of the composition descriptor based spectrum predictor}

\begin{figure}[h]
	\centering
	\begin{subfigure}{.45\textwidth}
		\includegraphics[width=\textwidth]{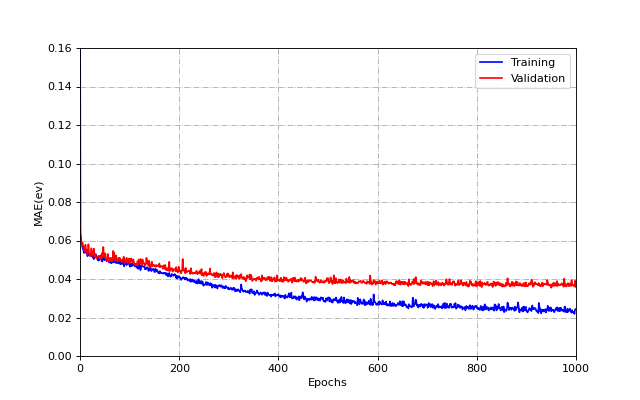}
		\caption{MAE changes during training}
		\vspace{3pt}
	\end{subfigure}
	\begin{subfigure}{.45\textwidth}
		\includegraphics[width=\textwidth]{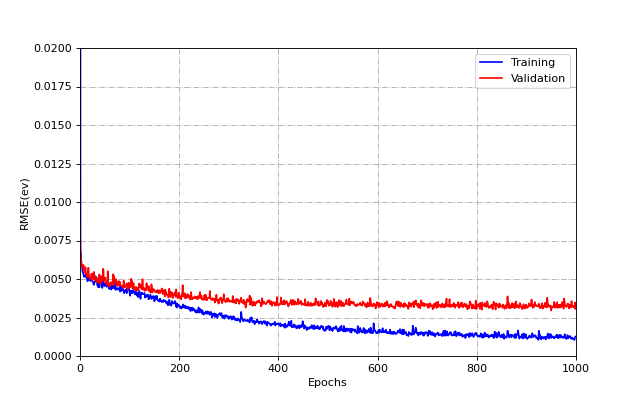}
		\caption{RMSE changes during training}
		\vspace{3pt}
	\end{subfigure}
	\begin{subfigure}{.45\textwidth}
		\includegraphics[width=\textwidth]{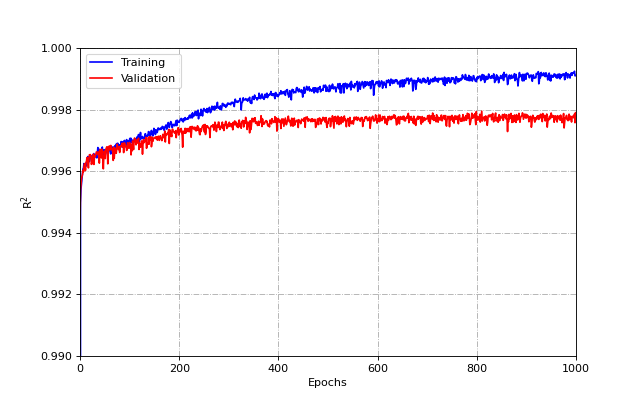}
		\caption{R$^2$ changes during training}
		\vspace{3pt}
	\end{subfigure}	
	\caption{Training and validation errors during training for composition based spectrum prediction.}
    \label{fig:prediction_performance}
\end{figure}

Considering the small sample data for a specific element set, we first train a fully connected neural network Model 1 on Dataset A, and then use the transfer learning method to fine tuning the model parameters of Model 2 whose initial parameters are transferred from Model 1 using Dataset B. Figure \ref{fig:prediction_performance} shows how the training and validation performance criteria MAE, RMSE and R$^2$ change during Model 1 training. After 800 epochs the MAE and RMSE are reduced to 0.04ev and 0.004ev and R$^2$ is raised to 0.997. This model will be used as the source model to build target prediction models for different chemical systems of specific element sets.

After training the source Model 1, we transfer the model parameters to the target chemical system with a given set of elements and fine-tune its parameters using the samples of the specified chemical system. We calculate the performance of the prediction models for target formula with few samples and with many samples to compare the performances of the models trained with or without using transfer learning strategy. In the process of training, we use randomly selected target formula as the test set. In transfer learning progress, the Dataset B is independently divided into training set, validation set and test set (only the target formula). The results are shown in Table \ref{table:transfer-learning}. For formula group [Sn, Ca, Zr, Hf] and [Fe, Bi, V, Mn] consisting of 88 and 973 different formula samples respectively, we randomly choose two quaternary compounds $Sn\textsubscript{0.3}Ca\textsubscript{0.1}Zr\textsubscript{0.4}Hf\textsubscript{0.2}$ and $Fe\textsubscript{0.25}Bi\textsubscript{0.15}V\textsubscript{0.3}Mn\textsubscript{0.3}$ as representative test formulas with small number of samples and with large large number of samples, respectively.

\begin{table}[h]
\begin{center}
\caption{Testing errors for composition based spectrum prediction.}
\label{table:transfer-learning}
\begin{tabular}{l|lll}
\hline
                        & R$^2$     & MAE (ev) & RMSE (ev) \\ \hline
small sample set without TL & 0.9925 & 0.0936   & 0.0221    \\ \hline
small sample set with TL    & 0.9947 & 0.0682   & 0.0127    \\ \hline
large sample set without TL & 0.9953 & 0.0416   & 0.0033    \\ \hline
large sample set with TL    & 0.9982 & 0.0251   & 0.0012  \\ \hline
\end{tabular}
\end{center}
\end{table}

\subsection{Comparison of two GA encoding-decoding methods}

\begin{figure}[ht]
	\centering
	\begin{subfigure}{.45\textwidth}
		\includegraphics[width=\textwidth]{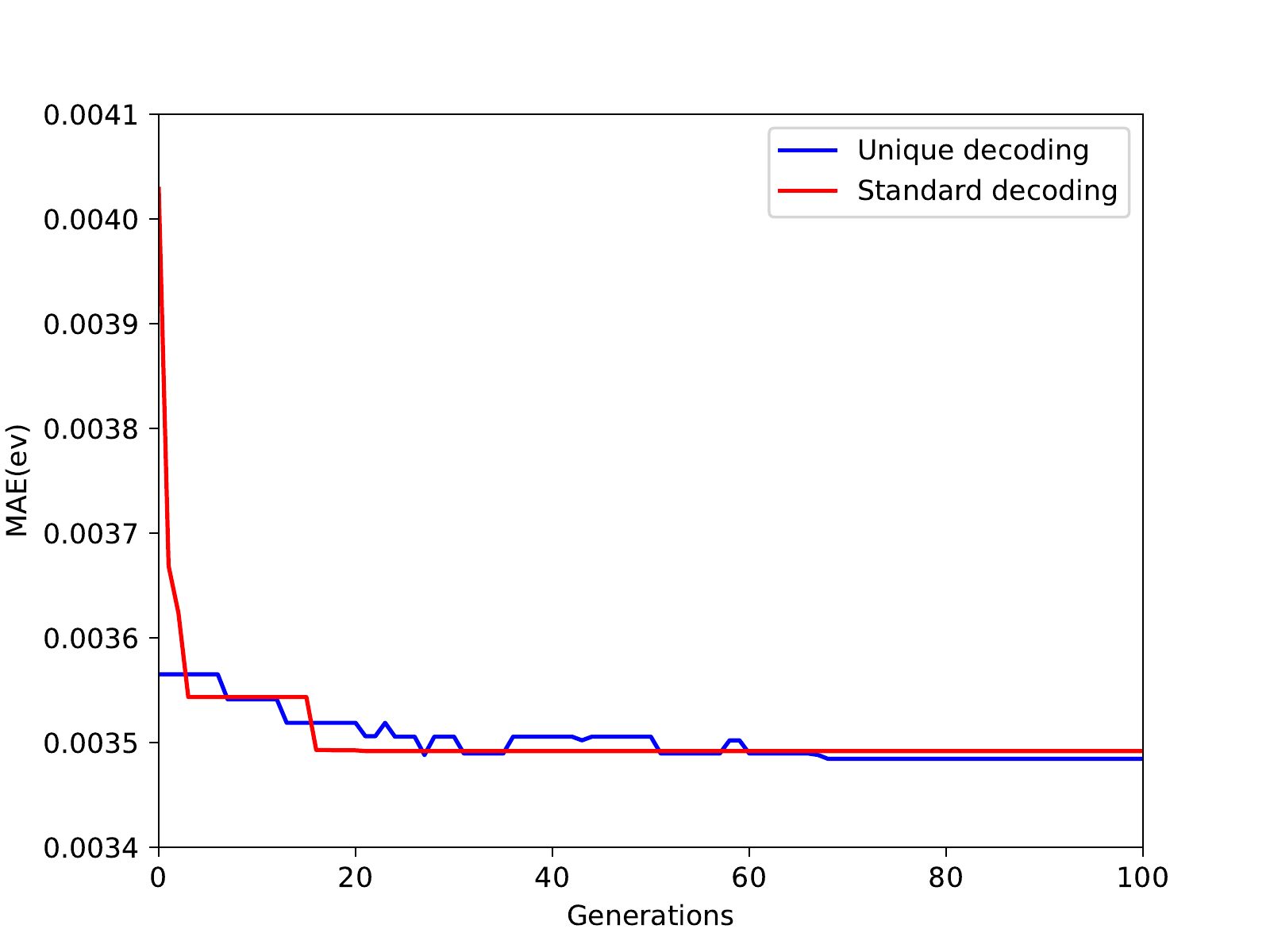}
		\caption{MAEs of  $Ni\textsubscript{0.15}Bi\textsubscript{0.7}Mn\textsubscript{0.15}$.}
		\vspace{3pt}
	\end{subfigure}
	\begin{subfigure}{.45\textwidth}
		\includegraphics[width=\textwidth]{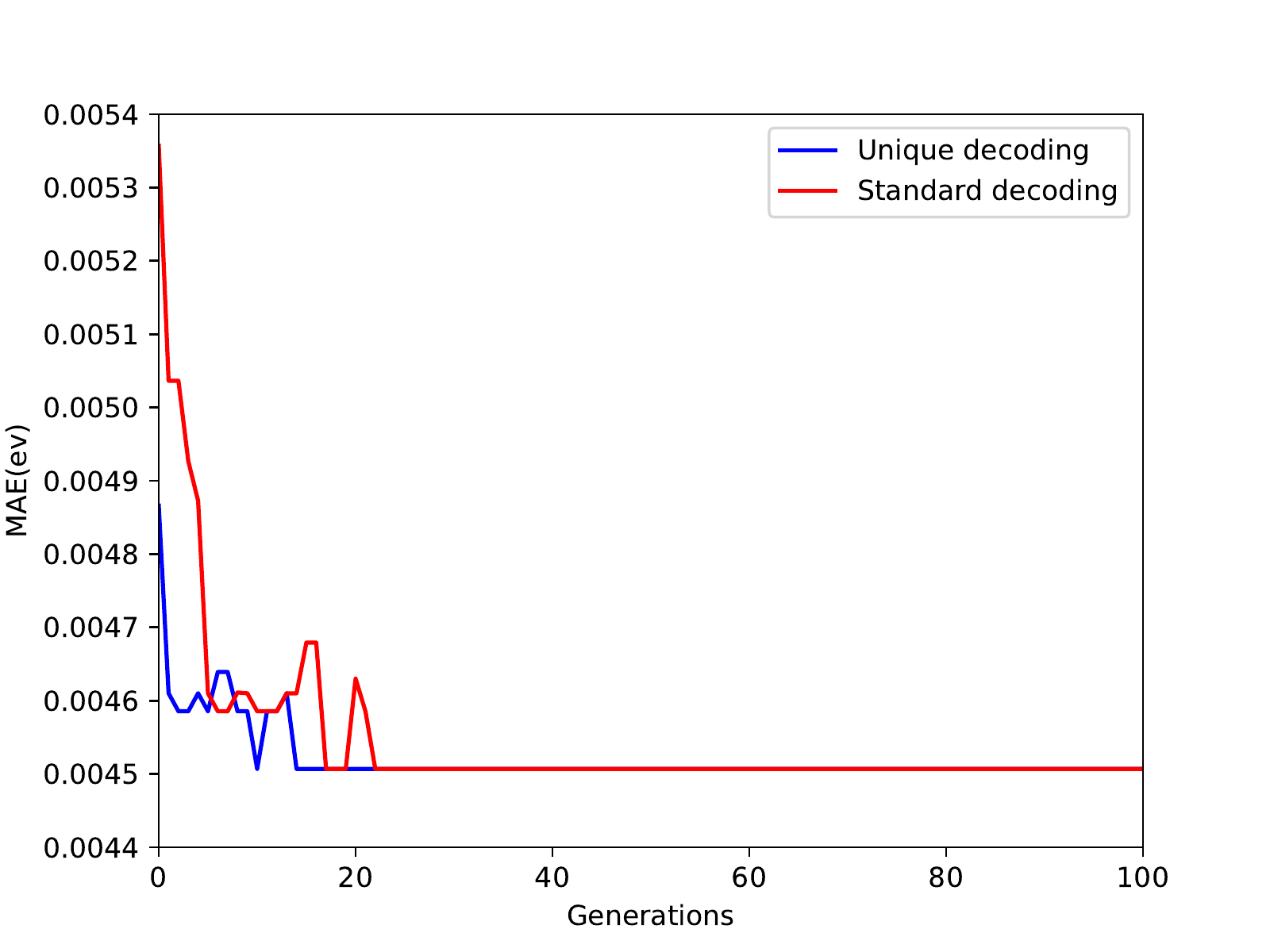}
		\caption{MAEs of  $Fe\textsubscript{0.05}V\textsubscript{0.05}Cu\textsubscript{0.25}Ca\textsubscript{0.65}$.}
		\vspace{3pt}
	\end{subfigure}
	\caption{MAE distances of two GA decoding methods.}
	\label{fig:compare encodings}
\end{figure}

To verify whether our proposed unique decoding approach in the GA for inverse design is better than the standard redundant decoding, we evaluate their performance on two inverse design problems of $Ni\textsubscript{0.25}Bi\textsubscript{0.67}Mn\textsubscript{0.08}$ and $Fe\textsubscript{0.05}V\textsubscript{0.05}Cu\textsubscript{0.25}Ca\textsubscript{0.65}$ respectively. Both experiments have been run with  30,000 evaluations with pop size of 300 and generation number 100. The MAE distances of our unique decoding (blue color) and standard redundant decoding (red color) are shown in Figure \ref{fig:compare encodings}. 

Through Figure \ref{fig:compare encodings}(a) we found that our decoding approach first achieved a lower MAE distance of 0.003488ev in the 27th generation and reached the minimum MAE 0.003484ev in the 68th generation, while MAE of the traditional decoding method converged to 0.003492ev in the 17th generation. As presented in Figure \ref{fig:compare encodings}(b), the minimum MAE of both decoding approaches reached 0.004507ev in 15th generation vs 23rd generation respectively, and our method converged faster. 

Both experiments proved that our unique decoding approach performed better than the standard decoding, not only reducing the redundant search space but also achieving the smallest MAE faster. 

\subsection{Inverse design of two compounds  using GA and BO}




To evaluate the inverse design performance of our TLOpt algorithm, we select two metal oxide materials from the whole dataset as the design target spectra with known composition information. Since most of the materials in the dataset are ternary and quaternary compounds, we randomly selected one for ternary $Ni\textsubscript{0.15}Bi\textsubscript{0.7}Mn\textsubscript{0.15}$ and one for quaternary compounds $
Fe\textsubscript{0.05}V\textsubscript{0.05}Cu\textsubscript{0.25}Ca\textsubscript{0.65}$ respectively.

\begin{figure}[ht!]
  \centering
  \includegraphics[width=0.85\linewidth]{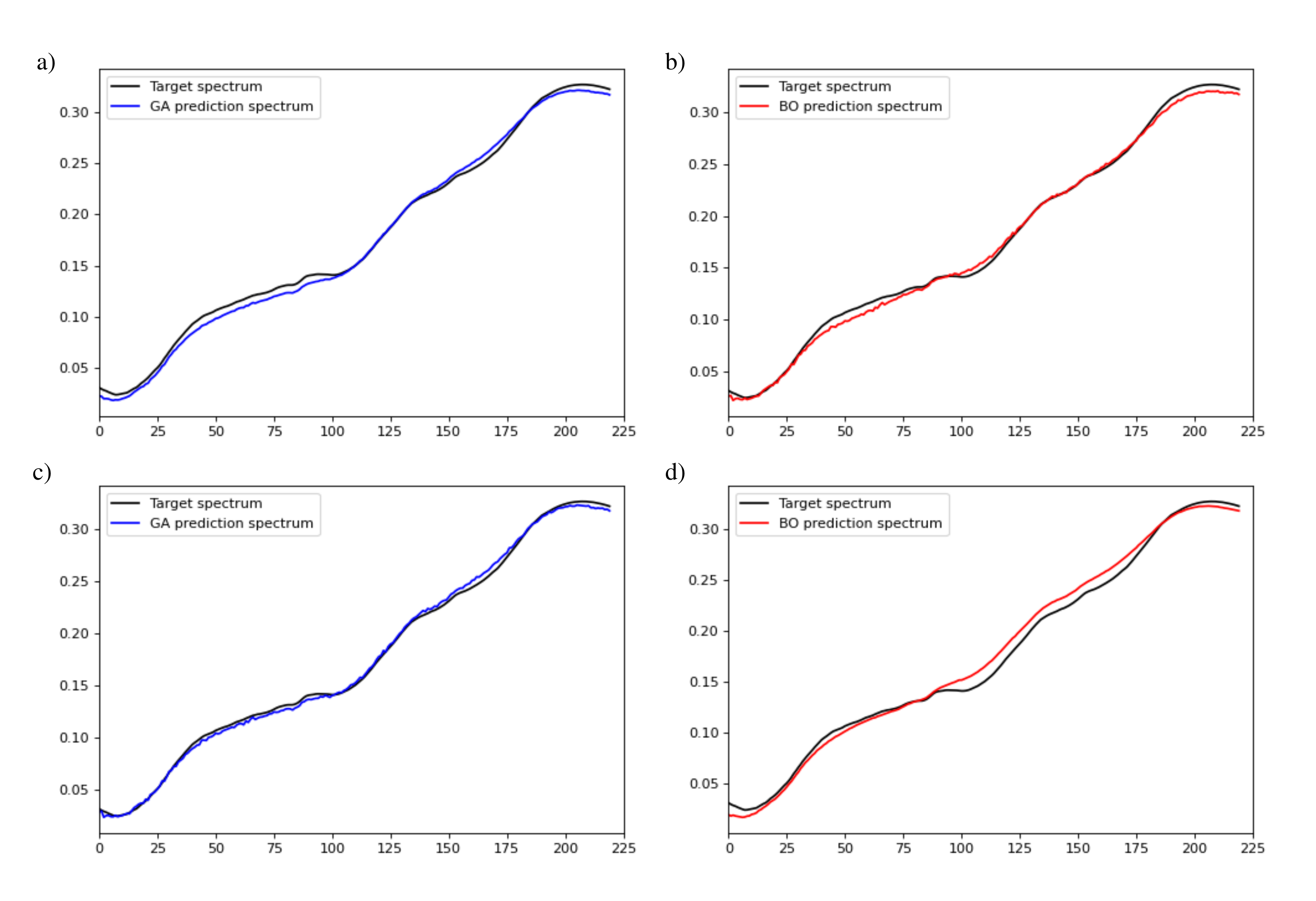}
  \caption{Inverse designs for target spectrum of $Ni\textsubscript{0.15}Bi\textsubscript{0.7}Mn\textsubscript{0.15}$. (a) Spectra comparison of inverse designs by GA with specified elements. (b) Spectra  comparison of inverse designs by BO with specified elements. (c) Spectra comparison of the inverse designs by GA without specifying composition elements. (d) Spectra comparison of inverse designs by BO without specifying composition elements}
  \label{figure6}
\end{figure}

\begin{figure}[h!]
  \centering
  \includegraphics[width=0.85\linewidth]{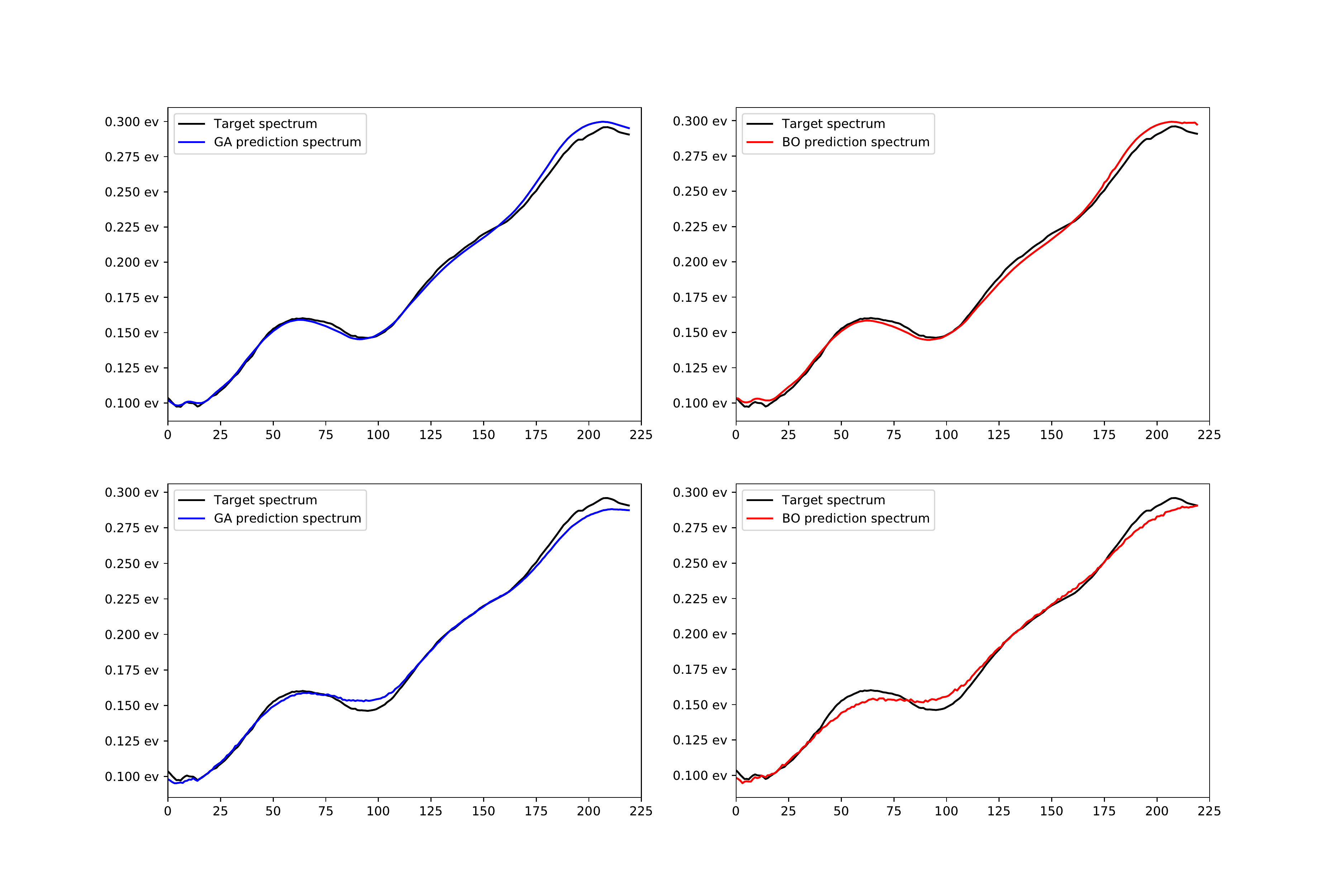}
  \caption{Inverse designed spectra of $Fe\textsubscript{0.05}V\textsubscript{0.05} Cu\textsubscript{0.25}Ca\textsubscript{0.65}$. (a) Spectra comparison of inverse designs by GA with specified elements. (b) Spectra comparison of inverse designs by BO with specified elements. (c) Spectra comparison of the inverse designs by GA without specifying composition elements. (d) Spectra comparison of inverse designs by BO without specifying composition elements}
  \label{figure7}
\end{figure}

For each target spectra, we conduct two inverse design tasks: one with specified element set so the inverse design algorithm only needs to determine the mole ratios of the given elements; in the second design task, the elements are not given so the inverse design algorithm also needs to search the elements along with the mole ratios of all elements. For both experiments, we tested our GA and BO-based algorithms based on the same number of evaluations to compare their performance. The final prediction results are shown in Figure \ref{figure6} and Figure \ref{figure7}.

 
 Figure\ref{figure6} (a) shows the true spectrum and predicted spectrum by the GA  search algorithm with composition elements [Ni, Bi, Mn] specified. After 100 generations, GA identified a formula $Ni\textsubscript{0.25}Bi\textsubscript{0.67}Mn\textsubscript{0.08}$ with very similar spectrum to the target spectrum. The MAE error of the predicted spectrum of GA is only 0.00274ev (See Table \ref{tab:predictedmaterials}). Figure \ref{figure6} (b) shows the spectrum of inverse designed metal oxide $Ni\textsubscript{0.04}Bi\textsubscript{0.71}Mn\textsubscript{0.25}$, which has a slightly higher MAE error 0.00278ev. Both spectra of the inverse design materials closely approximate the target spectra, which demonstrates the effectiveness of the proposed approach. Figure \ref{figure6} (c), (d) compare the prediction spectra and the target spectra of the inverse design metal oxides by GA and BO when the composition elements are not specified. After 100 generations, GA got the most similar spectrum with formula $Mn\textsubscript{0.22}Fe\textsubscript{0.45}Pd\textsubscript{0.33}$, and BO got $Gd\textsubscript{0.06}Ni\textsubscript{0.41}Fe\textsubscript{0.53}$. The MAEs of GA and BO are 0.00273ev and 0.00275ev respectively.

 Figure\ref{figure7} (a), (b) show the true spectrum and prediction spectrum of $Fe\textsubscript{0.05}V\textsubscript{0.05} Cu\textsubscript{0.25}Ca\textsubscript{0.65}$ by GA and BO with given composition elements [Fe, V, Cu, Ca]. After 100 generations, GA got the most similar spectrum with formula $Fe\textsubscript{0.08}V\textsubscript{0.07}Cu\textsubscript{0.25}Ca\textsubscript{0.6}$, and BO got $Fe\textsubscript{0.02}V\textsubscript{0.02}Cu\textsubscript{0.25}Ca\textsubscript{0.71}$. The MAEs of GA and BO are 0.00280ev and 0.00284ev respectively. Figure \ref{figure7} (c),(d) compare the prediction spectra and truly spectra of target formula by GA and BO when the composition elements are not specified. After 100 generations, GA got the most similar spectrum with formula $Al\textsubscript{0.15}Ti\textsubscript{0.47}Ba\textsubscript{0.11}Sc\textsubscript{0.27}$, and BO got $Mn\textsubscript{0.36}Yb\textsubscript{0.34}Mg\textsubscript{0.22}Pb\textsubscript{0.08}$. The MAEs of GA and BO are 0.00278ev and 0.00283ev respectively. The results are also shown in Table \ref{tab:predictedmaterials}.

\begin{figure}[ht]
  \centering
  \includegraphics[width=0.85\linewidth]{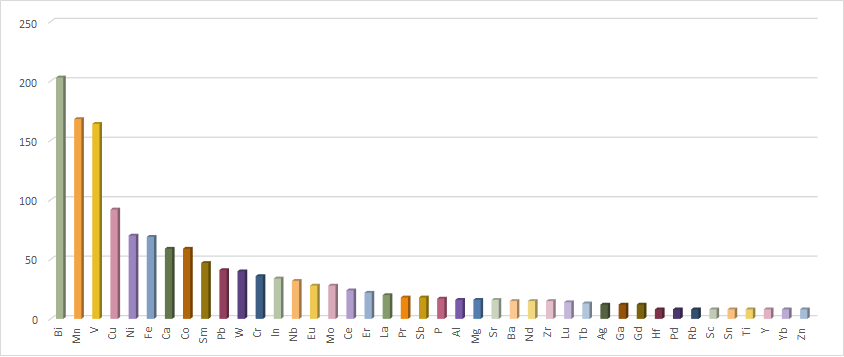}
  \caption{Frequency of all elements in the dataset.}
  \label{figure8}
\end{figure}

From Table\ref{tab:predictedmaterials} we can find that, for the same formula, GA performs better than BO method. When the composition elements are not specified, the prediction spectrum is closer to the true spectrum. For ternary and quaternary compounds, with same generations, the ternary compounds have better prediction performance. 
In the case of the same number of evaluations, BO requires much less time than GA, the time required for BO is about one-quarter to one-third of GA, but it is easy to fall into local optimization. GA can find the global optimum due to its characteristics of crossover, mutation and elite reservation, but it costs much more time. As the search space becomes larger and more combinations can be made, both GA and BO methods perform better when using random elements. And ternary compounds $Ni\textsubscript{0.15}Bi\textsubscript{0.7}Mn\textsubscript{0.15}$ perform better than quaternary $Fe\textsubscript{0.05}V\textsubscript{0.05}Cu\textsubscript{0.25}Ca\textsubscript{0.65}$ is determined by the characteristics of the dataset, the element Bi, Mn in [Ni, Bi, Mn] appear more frequently in the dataset, while the Fe, Cu, Ca in [Fe, V, Cu, Ca] appears relatively less frequently. The frequency of all elements in the dataset is shown as Figure \ref{figure8}.

\begin{table}[h!]
\caption{Predicted formulas and MAEs of experiment 1
}
\begin{center}
\begin{tabular}{l|ll|ll}
\hline
 Target->& \multicolumn{2}{c|}{Ni\textsubscript{0.15}Bi\textsubscript{0.7}Mn\textsubscript{0.15}} & \multicolumn{2}{c}{Fe\textsubscript{0.05}V\textsubscript{0.05}Cu\textsubscript{0.25}Ca\textsubscript{0.65}}\\\hline
Algorithm & Predicted Formula  & MAE(ev) &  Predicted Formula  & MAE(ev) \\\hline
GA with given  &  Ni\textsubscript{0.25}Bi\textsubscript{0.67}Mn\textsubscript{0.08} & 0.00274 & Fe\textsubscript{0.08}V\textsubscript{0.07}Cu\textsubscript{0.25}Ca\textsubscript{0.6} & 0.00280 \\
BO with given & Ni\textsubscript{0.04}Bi\textsubscript{0.71}Mn\textsubscript{0.25} & 0.00278 & Fe\textsubscript{0.02}V\textsubscript{0.02}Cu\textsubscript{0.25}Ca\textsubscript{0.71} & 0.00284 \\
GA with random & Mn\textsubscript{0.22}Fe\textsubscript{0.45}Pd\textsubscript{0.33} & 0.00273 & Al\textsubscript{0.15}Ti\textsubscript{0.47}Ba\textsubscript{0.11}Sc\textsubscript{0.27} & 0.00278\\
BO with random & Gd\textsubscript{0.06}Ni\textsubscript{0.41}Fe\textsubscript{0.53} & 0.00275 &Mn\textsubscript{0.36}Yb\textsubscript{0.34}Mg\textsubscript{0.22}Pb\textsubscript{0.08} & 0.00283\\
\hline
\end{tabular}
\end{center}
\label{tab:predictedmaterials}
\end{table}

\subsection{Prediction of the spectrum for a span of representative samples }

\begin{figure}[ht]
  \centering
  \includegraphics[width=0.9\linewidth]{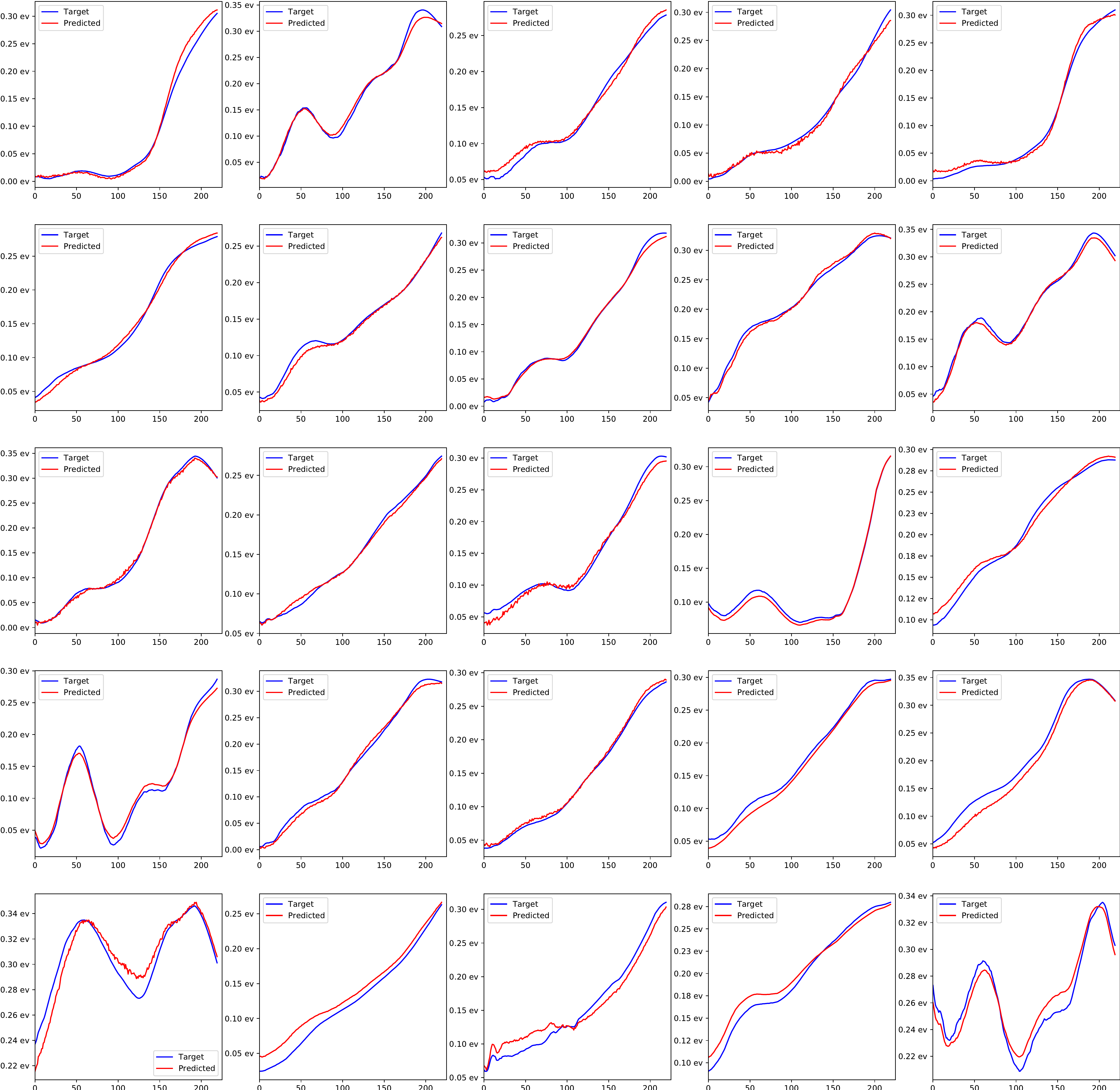}
  \caption{The inverse design of 25 metal oxides given the target spectra}
  \label{fig:figure9}
\end{figure}

To further evaluate the performance of our TLOpt inverse design algorithm, we randomly select 25 target compositions and their spectra as design targets from the whole 554 groups. The spectra (red color) of the inverse designed metal oxides predicted by the BO algorithm are shown in Figure \ref{fig:figure9} together with the target spectra (blue color).

From Figure\ref{fig:figure9}, we can find that most reconstructed spectra patterns (from row 1 to row 4) contain not only the general shape of the truth spectra but also finer details such as the presence of local maxima in absorption. For the targets in the last row in Figure \ref{fig:figure9}, the performance of the inverse design is still great, though not as good as those in top 4 rows. After close examination, we find that this is due to the composition elements Lu, Gd, Pd of these target materials rarely appear in the whole dataset while our TLOpt algorithm achieves great performance for the targets in the top 3 rows, which contain common composition elements Bi, Mn, V, Cu, and Ni. The results in Figure \ref{fig:figure9} thus validate our inverse design method on the dataset, proving the universality of the model. This inverse design model enables us to exploit hidden relationship between materials composition and their optical absorption properties and discover new materials with desired optical property.

\section{Conclusion}

We propose a transfer learning and global optimization based framework for inverse design of optical materials composition to achieve the target optical absorption spectrum. Our framework is composed of a fully connected neural network-based transfer learning model trained with Magpie features and global optimization based search including genetic algorithm and a Bayesian optimization model. Our transfer learning algorithm can be used to address the small dataset problem typical in material informatics, enabling our DNN model for predicting the material’s full UV-vis absorption spectrum from only its compound formula. Experiment of our transfer learning shows that after initial training and fine-tuning parameters through transfer learning, our prediction model performs well in spectrum prediction of metal oxide materials with only composition information alone. Extensive experiments show that our frame is able to discover interesting material compositions that approximate the target optical absorption spectrum. Our experiments also present that when running time is not an issue, genetic algorithm methods perform better than the Bayesian optimization in global optimization for our inverse design. Our research proves that machine learning based inverse design method could be used with small dataset when there are additional data of the same type used to initial training prediction model. Our inverse design model can be further improved when combined with materials structure information. Based on the successful design case studies, we believe our inverse design model is of great significance to be used to guide the discovery of new materials with other properties as well.

\section{Contribution}
Conceptualization, J.H.; methodology, J.H. and R.D.; software, R.D. and J.H.; validation, R.D. and J.H.;  investigation, R.D., J.H., Y.D.; resources, J.H.; writing--original draft preparation, R.D. and J.H.; writing--review and editing, J.H; visualization, R.D.,Y.D., and X.L.; supervision, J.H.;  funding acquisition, J.H.

\section{Acknowledgement}
Research reported in this work was supported in part by NSF under grant and 1940099 and 1905775 and by NSF SC EPSCoR Program under award number (NSF Award OIA-1655740 and GEAR-CRP 19-GC02). The views, perspective, and content do not necessarily represent the official views of the SC EPSCoR Program nor those of the NSF. This work was also partially supported.

\bibliography{references}

\begin{thebibliography}{10}

\bibitem{sikam2019study}
Pornsawan Sikam, Pairot Moontragoon, Zoran Ikonic, Thanayut Kaewmaraya, and
  Prasit Thongbai.
\newblock The study of structural, morphological and optical properties of (al,
  ga)-doped zno: Dft and experimental approaches.
\newblock {\em Applied Surface Science}, 480:621--635, 2019.

\bibitem{khalid2019synthesis}
Muhammad Khalid, Malik~Aman Ullah, Muhammad Adeel, Muhammad~Usman Khan,
  Muhammad~Nawaz Tahir, and Ataualpa Albert~Carmo Braga.
\newblock Synthesis, crystal structure analysis, spectral ir, uv--vis, nmr
  assessments, electronic and nonlinear optical properties of potent quinoline
  based derivatives: interplay of experimental and dft study.
\newblock {\em Journal of Saudi Chemical Society}, 23(5):546--560, 2019.

\bibitem{rajan2005materials}
Krishna Rajan.
\newblock Materials informatics.
\newblock {\em Materials Today}, 8(10):38--45, 2005.

\bibitem{ward2017atomistic}
Logan Ward and Chris Wolverton.
\newblock Atomistic calculations and materials informatics: A review.
\newblock {\em Current Opinion in Solid State and Materials Science},
  21(3):167--176, 2017.

\bibitem{gomez2018automatic}
Rafael G{\'o}mez-Bombarelli, Jennifer~N Wei, David Duvenaud, Jos{\'e}~Miguel
  Hern{\'a}ndez-Lobato, Benjam{\'\i}n S{\'a}nchez-Lengeling, Dennis Sheberla,
  Jorge Aguilera-Iparraguirre, Timothy~D Hirzel, Ryan~P Adams, and Al{\'a}n
  Aspuru-Guzik.
\newblock Automatic chemical design using a data-driven continuous
  representation of molecules.
\newblock {\em ACS central science}, 4(2):268--276, 2018.

\bibitem{popova2018deep}
Mariya Popova, Olexandr Isayev, and Alexander Tropsha.
\newblock Deep reinforcement learning for de novo drug design.
\newblock {\em Science advances}, 4(7):eaap7885, 2018.

\bibitem{lu2018accelerated}
Shuaihua Lu, Qionghua Zhou, Yixin Ouyang, Yilv Guo, Qiang Li, and Jinlan Wang.
\newblock Accelerated discovery of stable lead-free hybrid organic-inorganic
  perovskites via machine learning.
\newblock {\em Nature communications}, 9(1):1--8, 2018.

\bibitem{collins2016materials}
Sean~P Collins, Thomas~D Daff, Sarah~S Piotrkowski, and Tom~K Woo.
\newblock Materials design by evolutionary optimization of functional groups in
  metal-organic frameworks.
\newblock {\em Science advances}, 2(11):e1600954, 2016.

\bibitem{zunger2018inverse}
Alex Zunger.
\newblock Inverse design in search of materials with target functionalities.
\newblock {\em Nature Reviews Chemistry}, 2(4):1--16, 2018.

\bibitem{sanchez2018inverse}
Benjamin Sanchez-Lengeling and Al{\'a}n Aspuru-Guzik.
\newblock Inverse molecular design using machine learning: Generative models
  for matter engineering.
\newblock {\em Science}, 361(6400):360--365, 2018.

\bibitem{ikeda1997new}
Yuichi Ikeda.
\newblock A new method of alloy design using a genetic algorithm and molecular
  dynamics simulation and its application to nickel-based superalloys.
\newblock {\em Materials transactions, JIM}, 38(9):771--779, 1997.

\bibitem{molesky2018inverse}
Sean Molesky, Zin Lin, Alexander~Y Piggott, Weiliang Jin, Jelena Vuckovi{\'c},
  and Alejandro~W Rodriguez.
\newblock Inverse design in nanophotonics.
\newblock {\em Nature Photonics}, 12(11):659--670, 2018.

\bibitem{piggott2017fabrication}
Alexander~Y Piggott, Jan Petykiewicz, Logan Su, and Jelena Vu{\v{c}}kovi{\'c}.
\newblock Fabrication-constrained nanophotonic inverse design.
\newblock {\em Scientific reports}, 7(1):1--7, 2017.

\bibitem{liu2018training}
Dianjing Liu, Yixuan Tan, Erfan Khoram, and Zongfu Yu.
\newblock Training deep neural networks for the inverse design of nanophotonic
  structures.
\newblock {\em ACS Photonics}, 5(4):1365--1369, 2018.

\bibitem{peurifoy2018nanophotonic}
John Peurifoy, Yichen Shen, Li~Jing, Yi~Yang, Fidel Cano-Renteria, Brendan~G
  DeLacy, John~D Joannopoulos, Max Tegmark, and Marin Solja{\v{c}}i{\'c}.
\newblock Nanophotonic particle simulation and inverse design using artificial
  neural networks.
\newblock {\em Science advances}, 4(6):eaar4206, 2018.

\bibitem{jiang2019simulator}
Jiaqi Jiang and Jonathan~A Fan.
\newblock Simulator-based training of generative neural networks for the
  inverse design of metasurfaces.
\newblock {\em Nanophotonics}, 1(ahead-of-print), 2019.

\bibitem{liu2018generative}
Zhaocheng Liu, Dayu Zhu, Sean~P Rodrigues, Kyu-Tae Lee, and Wenshan Cai.
\newblock Generative model for the inverse design of metasurfaces.
\newblock {\em Nano letters}, 18(10):6570--6576, 2018.

\bibitem{pestourie2018inverse}
Rapha{\"e}l Pestourie, Carlos P{\'e}rez-Arancibia, Zin Lin, Wonseok Shin,
  Federico Capasso, and Steven~G Johnson.
\newblock Inverse design of large-area metasurfaces.
\newblock {\em Optics express}, 26(26):33732--33747, 2018.

\bibitem{aharoni2018universal}
Hillel Aharoni, Yu~Xia, Xinyue Zhang, Randall~D Kamien, and Shu Yang.
\newblock Universal inverse design of surfaces with thin nematic elastomer
  sheets.
\newblock {\em Proceedings of the National Academy of Sciences},
  115(28):7206--7211, 2018.

\bibitem{freeze2019search}
Jessica~G Freeze, H~Ray Kelly, and Victor~S Batista.
\newblock Search for catalysts by inverse design: artificial intelligence,
  mountain climbers, and alchemists.
\newblock {\em Chemical reviews}, 119(11):6595--6612, 2019.

\bibitem{sanchez2017optimizing}
Benjamin Sanchez-Lengeling, Carlos Outeiral, Gabriel~L Guimaraes, and Al{\'a}n
  Aspuru-Guzik.
\newblock Optimizing distributions over molecular space. an
  objective-reinforced generative adversarial network for inverse-design
  chemistry (organic).
\newblock 2017.

\bibitem{doscher2014sunlight}
H~D{\"o}scher, JF~Geisz, TG~Deutsch, and JA~Turner.
\newblock Sunlight absorption in water--efficiency and design implications for
  photoelectrochemical devices.
\newblock {\em Energy \& Environmental Science}, 7(9):2951--2956, 2014.

\bibitem{liao2020metaheuristic}
T~Warren Liao and Guoqiang Li.
\newblock Metaheuristic-based inverse design of materials--a survey.
\newblock {\em Journal of Materiomics}, 2020.

\bibitem{qin2020genetic}
Longhui Qin, Weicheng Huang, Yayun Du, Luocheng Zheng, and Mohammad~Khalid
  Jawed.
\newblock Genetic algorithm-based inverse design of elastic gridshells.
\newblock {\em Structural and Multidisciplinary Optimization}, pages 1--17,
  2020.

\bibitem{li2017bayesian}
Xiaoyang Li, Yuqing Hu, Enrico Zio, and Rui Kang.
\newblock A bayesian optimal design for accelerated degradation testing based
  on the inverse gaussian process.
\newblock {\em IEEE Access}, 5:5690--5701, 2017.

\bibitem{khadilkar2017inverse}
Mihir~R Khadilkar, Sean Paradiso, Kris~T Delaney, and Glenn~H Fredrickson.
\newblock Inverse design of bulk morphologies in multiblock polymers using
  particle swarm optimization.
\newblock {\em Macromolecules}, 50(17):6702--6709, 2017.

\bibitem{zhang2015inverse}
Yue-Yu Zhang, Weiguo Gao, Shiyou Chen, Hongjun Xiang, and Xin-Gao Gong.
\newblock Inverse design of materials by multi-objective differential
  evolution.
\newblock {\em Computational Materials Science}, 98:51--55, 2015.

\bibitem{bureerat2018inverse}
Sujin Bureerat and Nantiwat Pholdee.
\newblock Inverse problem based differential evolution for efficient structural
  health monitoring of trusses.
\newblock {\em Applied Soft Computing}, 66:462--472, 2018.

\bibitem{wu2019learning}
Ziku Wu, Chang Ding, Guofeng Li, Xiaoming Han, and Juan Li.
\newblock Learning solutions to the source inverse problem of wave equations
  using ls-svm.
\newblock {\em Journal of Inverse and Ill-posed Problems}, 27(5):657--669,
  2019.

\bibitem{wirkert2016robust}
Sebastian~J Wirkert, Hannes Kenngott, Benjamin Mayer, Patrick Mietkowski,
  Martin Wagner, Peter Sauer, Neil~T Clancy, Daniel~S Elson, and Lena
  Maier-Hein.
\newblock Robust near real-time estimation of physiological parameters from
  megapixel multispectral images with inverse monte carlo and random forest
  regression.
\newblock {\em International journal of computer assisted radiology and
  surgery}, 11(6):909--917, 2016.

\bibitem{sun2015artificial}
Gang Sun, Yanjie Sun, and Shuyue Wang.
\newblock Artificial neural network based inverse design: Airfoils and wings.
\newblock {\em Aerospace Science and Technology}, 42:415--428, 2015.

\bibitem{stein2019machine}
Helge~S Stein, Dan Guevarra, Paul~F Newhouse, Edwin Soedarmadji, and John~M
  Gregoire.
\newblock Machine learning of optical properties of materials--predicting
  spectra from images and images from spectra.
\newblock {\em Chemical science}, 10(1):47--55, 2019.

\bibitem{stein2019synthesis}
Helge~S Stein, Edwin Soedarmadji, Paul~F Newhouse, Dan Guevarra, and John~M
  Gregoire.
\newblock Synthesis, optical imaging, and absorption spectroscopy data for
  179072 metal oxides.
\newblock {\em Scientific data}, 6(1):1--5, 2019.

\bibitem{yu2013inverse}
Liping Yu, Robert~S Kokenyesi, Douglas~A Keszler, and Alex Zunger.
\newblock Inverse design of high absorption thin-film photovoltaic materials.
\newblock {\em Advanced Energy Materials}, 3(1):43--48, 2013.

\bibitem{dan2020generative}
Yabo Dan, Yong Zhao, Xiang Li, Shaobo Li, Ming Hu, and Jianjun Hu.
\newblock Generative adversarial networks (gan) based efficient sampling of
  chemical composition space for inverse design of inorganic materials.
\newblock {\em npj Computational Materials}, 6(1):1--7, 2020.

\bibitem{Stein2018Machine}
Helge~S. Stein, Guevarra Dan, Paul~F. Newhouse, Edwin Soedarmadji, and John~M.
  Gregoire.
\newblock Machine learning of optical properties of materials – predicting
  spectra from images and images from spectra.
\newblock {\em Chemical Ence}, 10, 2018.

\bibitem{ward2016general}
Logan Ward, Ankit Agrawal, Alok Choudhary, and Christopher Wolverton.
\newblock A general-purpose machine learning framework for predicting
  properties of inorganic materials.
\newblock {\em npj Computational Materials}, 2(1):1--7, 2016.

\bibitem{cao2019convolutional}
Zhuo Cao, Yabo Dan, Zheng Xiong, Chengcheng Niu, Xiang Li, Songrong Qian, and
  Jianjun Hu.
\newblock Convolutional neural networks for crystal material property
  prediction using hybrid orbital-field matrix and magpie descriptors.
\newblock {\em Crystals}, 9(4):191, 2019.

\bibitem{r1}
\url{https://reference.wolfram.com/language/note/ElementDataSourceInformation.html}.

\bibitem{zhuang2020comprehensive}
Fuzhen Zhuang, Zhiyuan Qi, Keyu Duan, Dongbo Xi, Yongchun Zhu, Hengshu Zhu, Hui
  Xiong, and Qing He.
\newblock A comprehensive survey on transfer learning.
\newblock {\em Proceedings of the IEEE}, 2020.

\bibitem{holland1975adaptation}
JohnH Holland.
\newblock adaptation in natural and artificial systems, university of michigan
  press, ann arbor,”.
\newblock {\em Cit{\'e} page}, 100, 1975.

\bibitem{huning1976evolutionsstrategie}
Alois Huning.
\newblock Evolutionsstrategie. optimierung technischer systeme nach prinzipien
  der biologischen evolution, 1976.

\bibitem{srinivas1994genetic}
Mandavilli Srinivas and Lalit~M Patnaik.
\newblock Genetic algorithms: A survey.
\newblock {\em computer}, 27(6):17--26, 1994.

\bibitem{pakhnova2020search}
Maria Pakhnova, Ivan Kruglov, Alexey Yanilkin, and Artem~R Oganov.
\newblock Search for stable cocrystals of energetic materials using the
  evolutionary algorithm uspex.
\newblock {\em Physical Chemistry Chemical Physics}, 2020.

\bibitem{jennings2019genetic}
Paul~C Jennings, Steen Lysgaard, Jens~Strabo Hummelsh{\o}j, Tejs Vegge, and
  Thomas Bligaard.
\newblock Genetic algorithms for computational materials discovery accelerated
  by machine learning.
\newblock {\em npj Computational Materials}, 5(1):1--6, 2019.

\bibitem{shariati2020prediction}
Mahdi Shariati, Mohammad~Saeed Mafipour, Peyman Mehrabi, Masoud Ahmadi, Karzan
  Wakil, Nguyen~Thoi Trung, and Ali Toghroli.
\newblock Prediction of concrete strength in presence of furnace slag and fly
  ash using hybrid ann-ga (artificial neural network-genetic algorithm).
\newblock {\em Smart Structures and Systems}, 25(2):183--195, 2020.

\bibitem{Gobin2008On}
Oliver~C. Gobin and Ferdi Schüth.
\newblock On the suitability of different representations of solid catalysts
  for combinatorial library design by genetic algorithms.
\newblock {\em Journal of Combinatorial Chemistry}, 10(6):835--846, 2008.

\bibitem{kushner1964new}
Harold~J Kushner.
\newblock A new method of locating the maximum point of an arbitrary multipeak
  curve in the presence of noise.
\newblock 1964.

\bibitem{jones1998efficient}
Donald~R Jones, Matthias Schonlau, and William~J Welch.
\newblock Efficient global optimization of expensive black-box functions.
\newblock {\em Journal of Global optimization}, 13(4):455--492, 1998.

\bibitem{snoek2012practical}
Jasper Snoek, Hugo Larochelle, and Ryan~P Adams.
\newblock Practical bayesian optimization of machine learning algorithms.
\newblock In {\em Advances in neural information processing systems}, pages
  2951--2959, 2012.

\end{thebibliography}
\bibliographystyle{unsrt}

\end{document}